\documentstyle[12pt,titlepage]{article}
\newif\ifnoepsf 
\newif\ifpublisher
\ifnoepsf\else\input{epsf.sty}\fi

\newcommand{\figinclude}[2]%
{\ifnoepsf\fbox{figure {\tt #2} goes here}%
\else{\epsfxsize=#1 pt \epsfbox{#2}}\fi}

\newcommand{\PHI}{%
  \unitlength=0.85cm
  \begin{picture}(1,1)
    \put(0,0){\makebox(1,1){$\phi$}}
  \end{picture}}
\newcommand{\V}{%
  \unitlength=0.85cm
  \begin{picture}(0.5,1)
    \put(0,0){\makebox(0.5,1){$v$}}
  \end{picture}}
  \newcommand{\PHIPM}{%
  \unitlength=0.85cm
  \begin{picture}(1,1)
    \put(0,0){\makebox(1,1){$\phi_{\pm}$}}
  \end{picture}}
\newcommand{\VPM}{%
  \unitlength=0.85cm
  \begin{picture}(0.5,1)
    \put(0,0){\makebox(0.5,1){$v_{\pm}$}}
  \end{picture}}
\newcommand{\ZERO}{%
  \unitlength=0.5cm
  \begin{picture}(1,1)
    \put(0,0){\makebox(1,1){{\huge 0}}}
  \end{picture}}
\newcommand{\PHIONE}{%
  \unitlength=0.5cm
  \begin{picture}(1,1)
    \put(0.1,0.5){$\phi_1$}
  \end{picture}}
\newcommand{\PHITWO}{%
  \unitlength=0.5cm
  \begin{picture}(1,1)
    \put(0.25,0.11){$\phi_{2^n}$}
  \end{picture}}
\newcommand{\Z}{{\sf Z\hspace*{-0.93ex}Z}\hspace*{0.1ex}}
\newcommand{\tls}{\la\lambda_{\rm s}\ra}
\newcommand{\too}{\rightarrow}
\newcommand{\mod}{{\rm mod}}
\newcommand{\OO}{O}
\newcommand{\Tr}{\mbox{Tr}}
\newcommand{\unity}{{\bf 1}}
\newcommand{\tas}{\la\alpha_{\rm s}\ra}
\newcommand{\as}{\alpha_{\rm s}}
\newcommand{\asa}[1]{\alpha_{{\rm s},#1}}

\newcommand{\bs}{\mbox{\boldmath $\sigma$}}
\newcommand{\bp}{\mbox{\boldmath $\phi$}}
\newcommand{\phip}{\phi_+}
\newcommand{\phim}{\phi_-}
\newcommand{\phipm}{\phi_\pm}
\newcommand{\awa}[1]{\la\hat{W}_{#1}(z)\ra}
\newcommand{\gc}{g_*}
\newcommand{\la}{\left\langle}
\newcommand{\ra}{\right\rangle}

\renewcommand{\theequation}{\arabic{section}.\arabic{equation}}
\newcommand{\beq}{\begin{equation}}
\newcommand{\eeq}{\end{equation}}
\newcommand{\bea}{\begin{eqnarray}}
\newcommand{\eea}{\end{eqnarray}}
\newcommand{\tr}{\, {\rm tr} \,}
\newcommand{\e}{\, {\rm e}}
\newcommand{\E}{{\rm e}}

\newcommand{\lt}{\left(}
\newcommand{\rt}{\right)}
\newcommand{\hf}{\frac{1}{2}}

\newcommand{\oon}{\frac{1}{N}}

\newcommand{\pa}{\partial}
\newcommand{\p}{\phi}

\newcommand{\bt}{{\boldmath \tau}}

\newcommand{\be}{\begin{equation}}
\newcommand{\ee}{\end{equation}}
\newcommand{\ba}{\begin{eqnarray}}
\newcommand{\ea}{\end{eqnarray}}
\renewcommand{\theequation}{\arabic{section}.\arabic{equation}}
\newcounter{subeqn}
\begin{document}
\begin{titlepage}
\addtolength{\baselineskip}{.7mm}
\thispagestyle{empty}
\begin{flushright}
TIT/HEP--261\\
NUP-A-94-16 \\
hep-th/9409009\\
September, 1994
\end{flushright}
\begin{center}
{\large{\bf  Renormalization Group Flow \\
in One- and Two-Matrix Models }} \\[15mm]
{\sc Saburo Higuchi}
\footnote{{\tt e-mail: hig@th.phys.titech.ac.jp}, JSPS fellow\\
\indent \ address after 7 September 1994: 
Service de Physique Th\'eorique,\\ 
\indent \ CEA Saclay, 91191 \mbox{Gif-sur-Yvette} CEDEX, France},\ \   
{\sc Chigak Itoi}
\footnote{\tt e-mail: itoi@phys.cst.nihon-u.ac.jp} ,\\
{\sc Shinsuke Nishigaki}
\footnote{{\tt e-mail: nsgk@th.phys.titech.ac.jp} \\
\indent \ address after 1 October 1994:  Department of Physics,\\
\indent \ Technion - Israel Institute of Technology, Haifa 32000, Israel} 
\ \ and \ \ 
{\sc Norisuke Sakai}
\vspace{10mm}
\footnote{\tt e-mail: nsakai@th.phys.titech.ac.jp} \\[2mm]
$^{{\small *,\ \ddag ,\ \S}}${\it Department of Physics,
Tokyo Institute of Technology, \\[1mm]
Oh-okayama, Meguro, Tokyo 152, Japan} \\ 
\vspace{5mm}
$^{{\small \dagger}}${\it Department of Physics, \\[1mm]
College of Science and Technology, Nihon University, \\[1mm]
Kanda Surugadai, Chiyoda, Tokyo 101, Japan} \\[6mm]
\vfill
{\bf Abstract}\\[5mm]
\end{center}
Large-$N$ renormalization group equations
for one- and two-matrix models are derived. 
The exact renormalization group equation involving
infinitely many induced interactions   
can be rewritten in a form that has a finite number of coupling 
constants by taking account of reparametrization identities. 
Despite the nonlinearity of the equation, the location 
of fixed points and the scaling exponents can be extracted 
from the equation. They agree with the spectrum 
of relevant operators in
the exact solution. 
A linearized $\beta$-function approximates well the global 
phase structure which includes several nontrivial fixed points. 
The global renormalization group flow suggests a kind of $c$-theorem 
in two-dimensional quantum gravity.

{\parbox{13cm}{\hspace{5mm}
}}

\vspace*{5ex}
\end{titlepage}
\setcounter{footnote}{0}
%%%%% Intro %%%%%%%%%%%%%%%%%%%%%%%%%%%%%
\section{Introduction} 
Two-dimensional quantum gravity is important to 
explore the nonperturbative dynamics of string models 
and statistical models of random surface as well as to
provide a theoretical laboratory for 
quantum gravity in higher dimensions.  
The conformal field theory approach has
successfully described two-dimensional 
quantum gravity and evaluated its universal quantities
\cite{KnPoZa:LightCone}.
However, the continuum description has encountered a
difficulty where quantum gravity is coupled to the conformal matter 
field with the central charge $c > 1$,
despite its importance furnishing off-critical string models
in an arbitrary dimensional target space.
 
As a discretized approach, the matrix model 
\cite{BrItPaZu:planar,BeItZu:qfttech} provides a more rigorous 
and powerful formulation of two-dimensional quantum gravity 
and off-critical string models.
Exact solutions in matrix models have been obtained so far only 
for cases corresponding to matter theories with $c \le 1$ coupled to 
quantum gravity \cite{KaKoMi:DT}--\cite{GrKl:MQM}.
It is straightforward to define matrix model candidates for 
quantum gravity coupled to conformal matter with $c > 1$, such as 
multi-Ising models \cite{BrHi:n-Ising1}. 
However, it has been difficult to solve these theories exactly. 
Therefore it is most desirable to find a reliable approximation method 
which gives correct results for solved models and reasonable results 
for so far unsolved models. 

Br\'{e}zin and Zinn-Justin have proposed a large-$N$ 
renormalization group (RG) equation as such an approximation method 
\cite{Carlson:RGN,BrZi:RGMat}.
Subsequently, we have explicitly derived exact RG 
equations for $O(N)$-vector models \cite{HiItSa:Vector,HiItSa:MV}
and one-matrix models \cite{HiItNiSa:Matrix}.
We found that it is crucial to take account of the reparametrization 
identities in order to obtain a meaningful RG 
equation. Moreover, the RG equation turns out to be a nonlinear 
equation for the matrix model in contrast to the linear one for vector 
models. 

The purpose of the present paper is to explain
our large-$N$ RG approach for matrix models in depth and 
to report on results for a two-matrix model. 
We obtain correct locations of several nontrivial
fixed points and scaling exponents. 
With this result, we can identify the matrix model representation 
of the gravitationally dressed operators corresponding to those in the 
continuum theory. 
Moreover, we find that the large-$N$ RG
is useful to describe the global phase structure of matrix models.
The global picture of the RG flow can be drawn practically 
by a linearized approximation for the nonlinear 
RG equation. 
It should be noted that the linearized 
approximation is meaningful only 
after the reparametrization identities are taken into account. 
We observe from this global flow that 
an RG trajectory connects a pair of fixed points corresponding to two 
unitary conformal matter theories
coupled to gravity,
in accordance with the expectation 
based on conformal field theories over fixed background 
\cite{Zamolodchikov:ctheorem}.
This flow can be understood by means of the usual 
Kadomtsev-Petviashvili (KP) flow 
\cite{Douglas:KdV}--\cite{FuKaNa:universal}.
On the other hand, we also find that a pair of fixed points are 
connected by an RG trajectory which 
is out of the KP hierarchical description. 
{From} these examples, we conjecture that a kind of the gravitational 
analogue of Zamolodchikov's $c$-theorem
holds in accordance with 
some previous expectations \cite{KuSa:gravctheorem}
for several cases.

The partition function $Z_N(g_j)$ of a one-matrix model with a
general potential $V(\phi)$ is defined by an integral over an
$N\times N$ hermitian matrix $\phi$
\begin{eqnarray}
  Z_N(g_j) & = &  \int d \phi \exp \left[-N  \tr V(\phi)\right],
  \label{eqn:mat-part} \\
  & & V(\phi) = \sum_{k\geq 1} \frac{g_k}{k} \phi^k, \nonumber
  \label{eqn:mat-generic-pot}
\end{eqnarray}
In particular, we are interested in the subspace
$\{g_1=0,\ g_2=1,\ g_j\equiv 0 
\mbox{\ for\ } j> {}^\exists m \}$. 
Let us define the free energy by 
\begin{equation}
  F(N,g_j) = 
  - \frac{1}{N^2} \log 
\left[\frac{Z_N(g_j)}{Z_N(g_2=1, \mbox{others}=0)}\right].
  \label{eqn:mat-free-energy}
\end{equation}
The free energy of the matrix model
gives a generating function of connected 
Feynman diagrams constructed in terms of 
$j$-point vertices with the coupling $g_j$.  
By considering the dual of the Feynman diagrams, 
we can interpret that 
the dual diagram gives a decomposition 
of a two-dimensional surface in terms of $j$-gons corresponding 
to the $j$-point vertices $g_j$. 
Under this interpretation,
the free energy of the matrix model is identified with the
partition function of the two-dimensional quantum gravity.
One finds that contributions 
from the different genera are distinguished by powers of $1/N^2$.
 
For brevity, we consider the triangulation 
$(g_3=g,\ g_j\equiv0 \mbox{\ for\ } j> 3) $.
The free energy $F^h (g)$ 
for genus $h$  surfaces is given by
the coefficient of $N^{-2h}$ in the $1/N$-expansion, 
and the number $n$ of triangles is counted by the power of $g$ 
\beq
  F(N,g)=\sum_{h=0}^{\infty} N^{-2h} F^h(g), \ \ 
   F^h(g)=\sum_{n} g^{n} 
   \nu_h(n).
\label{eqn:free-ener}
\eeq
Here $\nu_h(n)$ is the number of Feynman diagrams 
with $n$ vertices and genus $h$ 
divided by the symmetry factors. 
The free energy $F^h(g)$ can be regarded as
the partition function of the randomly triangulated surface
with genus $h$.
Its singular part behaves as
\begin{equation}
  F_{\rm sing}^h(g) \sim \left(g - g_* \right)^{-(\Gamma_{h}-2)} 
  + \mbox{less singular terms}
 \label{eqn:exact-sol}
\end{equation}
around a critical point $\gc$.
The singular behavior (\ref{eqn:exact-sol}) implies
that the number of triangles $\nu_h(n)$
has an asymptotic form
\begin{equation}
\nu_h (n) \sim {n}^{\Gamma_h - 3} g_*^{-n}\ \ \ 
(n \rightarrow \infty).
\end{equation}
The average number of triangles $\la n \ra$
of surfaces with genus $h$ is estimated by (\ref{eqn:exact-sol})
\begin{eqnarray}
\la n \ra  &=& \frac{ \sum_{n} n\ g^n 
 \ \nu_h(n)} {\sum_{n}  g^n 
\nu_h(n)} \nonumber \\
 &=& g\frac{\partial}{\partial g} \log F^h_{\rm sing} (g) \sim 
 {(\Gamma_h-2) g_* \over g_*-g}
 \rightarrow \infty 
 \quad \quad (g \rightarrow  g_* ). 
 \label{eqn:avr-nm}
\end{eqnarray}  
The average number of triangles 
in triangulations with any genus diverges
simultaneously at $g=g_*$ with the same exponent $-1$. 

In matrix models (and vector models \cite{Nishigaki:scalingvio})
solved so far,  
the susceptibility exponent $\Gamma_{h}$ depends linearly upon $h$ as 
\begin{equation}
  \Gamma_h = \gamma_0 + h \gamma_1 .
 \label{eqn:exponenth}
\end{equation}
Therefore we can take the
double scaling limit 
\begin{equation}
1/N \rightarrow 0,   \qquad g \rightarrow g_*,  \qquad 
{\rm with}   \quad   N^{-2/\gamma_1}(g - g_*)^{-1} 
\quad ={\rm fixed,}
\label{eqn:doublescalinglimit}
\end{equation}
and retain nontrivial contributions from all genera to the 
free energy.
In this limit the singular part of the free energy takes the form
\begin{equation}
  F_{\rm sing}(N,g) = (g-\gc)^{2-\gamma_0} f(N^{2/\gamma_1} (g-\gc)),
  \label{eqn:exact-sol-all-genera}
\end{equation}
where $f$ is an unspecified function.

Let us discuss the double scaling limit (\ref{eqn:doublescalinglimit}) 
from the viewpoint of 
the continuum limit in two-dimensional discretized quantum gravity. 
We consider the following
continuum limit for each genus $h$ by sending
the length $a$ of edges of 
triangles (spacing of the random lattice) to zero. 
If the average number of triangles $\la n \ra $ 
diverges, we can scale the length to zero 
so as to keep the `physical' area fixed:
\begin{equation}
        a \rightarrow 0, \quad 
        \la n \ra  \rightarrow \infty, \quad 
        \mbox{\  with\ } \quad 
        a^2 \cdot \la n \ra ={\rm fixed}.
\label{eqn:continuum-lim}
\end{equation}
Since the scaling behavior (\ref{eqn:avr-nm}) shows 
that the average number of triangles diverges by letting $g$ approach a 
critical value $g_*$, 
we can regard the double scaling limit 
(\ref{eqn:doublescalinglimit}) 
as the continuum limit eq.(\ref{eqn:continuum-lim}) 
for each genus $h$
\begin{equation}
  N^{-2/\gamma_1} \cdot  \la n \ra  
 =\mbox{fixed} \quad \quad \Leftrightarrow \quad \quad  
 a^2 \cdot \la n \ra  =\mbox{fixed}.
\label{eqn:cutoff}
\end{equation}
Therefore the double scaling limit suggests that 
$N^{-1/\gamma_1}$ is proportional to the lattice
spacing $a$. 
A similar point of view has been expressed previously
in refs.\cite{GrMi:DSL,Periwal}.

Br\'{e}zin and Zinn-Justin proposed the following approximation
scheme for calculating the location and the exponents of
critical points \cite{BrZi:RGMat}.
Integrating out parts of 
$(N+1)\times(N+1)$ components of the
matrix field $\phi$ in eq.(\ref{eqn:mat-part}) gives us the following
recursion relation 
in the first nontrivial order of perturbation:
\begin{equation}
  Z_{N+1}(g) = \lambda(N,g)^{N^2} Z_N(g+ \delta g(N,g)).
  \label{eqn:difference-formal}
\end{equation}
One expects that 
eq.(\ref{eqn:difference-formal}) is well approximated by a linear
differential equation
\begin{equation}
\left[  N \frac{\partial}{\partial N} - 
\beta(g) \frac{\partial}{\partial g} + \gamma (g) \right] F(N,g)
= r(g).
  \label{eqn:rge-formal}
\end{equation}
The homogeneous part (i.e.\ $r(g)$ set to zero) of 
eq.(\ref{eqn:rge-formal})
determines the singular part of $F$ 
around a zero $g=g_*$ (a fixed point) of the 
$\beta$-function as
\begin{equation}
  F_{\rm sing}(N,g)= (g - g_*)^{\gamma(g_*)/\beta'(g_*)} 
  f(N^{\beta'(g_*)} (g-g_*)) .
  \label{eqn:non-analytic-formal}
\end{equation}
This singular behavior is just the same as that 
of the exact solutions
(\ref{eqn:exact-sol-all-genera}) of the double scaled matrix model. 
Namely, the string susceptibility exponents are identified as
\begin{eqnarray}
   \Gamma_h 
& = &  \gamma_0 + h \gamma_1 \nonumber \\
& = &  \left(2 - \frac{\gamma(g_*)}{\beta'(g_*)}\right) 
  +  h \left(  \frac{2}{\beta'(g_*)}\right).
\label{eqn:mat-exponents}
\end{eqnarray}
Therefore once one knows the functions $\beta(g)$ and
$\gamma(g)$, one can compute positions and exponents of
critical points.
In fact, the RG equation computed in the one-loop approximation
provides us with a fairly good result \cite{BrZi:RGMat,AlDa}.
However we have confirmed in ref.\cite{HiItSa:MV} that 
no improvement is achieved if one pursues the 
perturbation to higher-loops by simply 
truncating induced interactions of higher orders. 
We now understand that the naive perturbative method fails 
due to the reparametrization invariance
of the partition function, as explained in sect.2. 

Eq.(\ref{eqn:rge-formal}) resembles
RG equations in field
theories in its form. 
Since eq.(\ref{eqn:cutoff}) shows that we can identify 
$N^{-1/\gamma_1}$ with the lattice
spacing $a$, we can regard an 
infinitesimal shift of $N$ as a substitute for a shift of the 
lattice spacing $a$.
Thus it may be  appropriate to call 
a differential equation describing 
the response of the physical quantity under the infinitesimal change of 
$N$, such as eq.(\ref{eqn:rge-formal}),
an RG equation for matrix models.

This paper is organized as follows.
In sect.2, we derive exact RG equations 
for one- and two-matrix models
via coset and/or eigenvalue methods. 
In sect.3, we develop a scheme to solve the RG 
equations and show that they reproduce the spectrum 
of the exact solutions.
In sect.4, we analyze the topology of the RG flow
using the linear approximation to the RG equations, and discuss
the gravitational $c$-theorem.
Possible applications of our method, including generalizations to
candidates for $c>1$ quantum gravity, are addressed in sect.5. 
The reparametrization identities for the two-matrix model are 
solved in appendix A. 

%%%%%%% Large N RQ Eq %%%%%%%%%%%%%%%%%%%%%
\section{Large-$N$ renormalization group equation}
\setcounter{equation}{0}

\subsection{One-matrix model I: coset method}

\noindent {\large \underline{RG transformation}}

In this subsection, we restrict ourselves to the simplest case $m=3$ and
set $g \equiv g_3 \geq 0$ without loss of generality.

To derive a relation (\ref{eqn:difference-formal}),
we employ a decomposition of
an $(N+1) \times (N+1)$ hermitian matrix variable $\Phi$
into an $N \times N$ hermitian matrix $\phi$,
a complex $N$-vector $v$,
its transposed conjugate $v^\dagger$
and a real scalar $\alpha$:
\begin{equation}
    \Phi = \left(
  \begin{array}{@{\,}c|c@{\,}}
  \PHI          & \V         \\ \hline
   v^\dagger & \alpha   \\
  \end{array}
  \right).
\end{equation} 
In terms of $\phi, v, v^\dagger$ and $\alpha$, the partition function
(\ref{eqn:mat-part}) reads
\beq
Z_{N+1} (g) = \int d\phi dv dv^\dagger d\alpha \ 
 \exp\left[ -(N+1) ( \tr V(\phi) + V(\alpha)+
      v^\dagger( \unity + g (\phi + \alpha \unity))v) \right].
  \label{eqn:decomp-part}
\eeq
First we integrate over $v$ and $v^\dagger$ exactly to obtain
\begin{eqnarray}
Z_{N+1}(g) & = &  
\left(\frac{\pi}{N+1}\right)^N
\int d\phi 
  \exp[ -(N+1) \tr V(\phi)] \nonumber \\
  & & \cdot \int d\alpha  
 \exp[ -(N+1) V(\alpha) - \tr \log (\unity + g (\phi +
  \alpha \unity)) ].
\end{eqnarray}
Now we are ready to evaluate the $\alpha$-integral 
by the saddle point method,
systematically in each order of the $1/N$-expansion.
The saddle point $\as=\as(g,\phi)$ is determined by the saddle point 
equation
\begin{equation}
  \as + g \as^2 + \frac gN
  \tr \frac{1}{\unity +  g(\phi+\as \unity)} = 0
  \label{eqn:spe}
\end{equation}
as a $U(N)$-invariant function in $\phi$
(i.e.\ a function in $\tr \phi^j$'s).

If we explicitly retain only the 
leading part of the $1/N$-expansion, 
we have 
\bea
\frac{Z_{N+1}(g)}{Z_N(g)} &=& \left(\frac{\pi}{N+1}\right)^N
\left\langle \exp[ -\tr V(\phi) - N V(\as) \right. \nonumber \\
& &  \left. -\tr \log( \unity + g(\phi + \as
\unity))   + \OO(N^0)] \right\rangle.
\label{prefact}
\eea
Here $\langle\cdots\rangle$ denotes the average 
$
Z_N^{-1} \int d\phi (\cdots) \exp [-N \tr V(\phi)] .
$
In the following we repeatedly utilize the factorization property 
of a multi-point function of $U(N)$-invariants  
${\cal O},\ {\cal O}'$ 
into a product of one-point functions,
\begin{equation}
  \langle {\cal O}\ {\cal O}' \rangle = 
  \langle {\cal O} \rangle \langle {\cal O}' \rangle + \OO(N^{-2})
\label{eqn:fact}
\end{equation}
in the large-$N$ limit.
In view of this factorization property, 
eq.(\ref{prefact}) can further be written as 
\bea
\frac{Z_{N+1}(g)}{Z_N(g)} 
& = & \left(\frac{\pi}{N+1}\right)^N
\left. \exp[ -\langle \tr V(\phi) \rangle - N V(\tas) \right. 
\nonumber \\ 
& &  \left. -\langle \tr \log( \unity + g(\phi + \tas
\unity)) \rangle  + \OO(N^0)] \right.  .
 \label{eqn:sp-difference}          
\end{eqnarray}
Finally we approximate a difference in
eq.(\ref{eqn:sp-difference}) by a differential to obtain
\begin{eqnarray}
\lefteqn{\left[ N\frac{\partial}{\partial N}+2 \right]  F(N,g)} 
  \label{eqn:sp-differential}
\\
 &\!\!\! =\!\!\! & - \frac 12 + \la \frac 1N  \tr V(\phi) \ra + 
V( \langle \as \rangle) +
\la \frac 1N \tr \log \left( \unity + g(\phi + \la \as \ra \unity) 
\right) \ra
  + O \lt \oon \rt.
  \nonumber
\end{eqnarray}
 
\noindent {\large \underline{Reparametrization invariance}}

The right hand side of eq.(\ref{eqn:sp-differential})
is a function of 
variables $N$, $g$ and 
$\la (1/N) \tr\phi^j\ra  = j(\partial F/\partial g_j) \ (j=1,2,\ldots)$.
What we will do here is to express 
the $\partial F/\partial g_j$-dependence $(j\neq 3)$
in terms of $\partial F/\partial g_3$.
This may be done by taking into account 
reparametrization invariance of the partition function 
(\ref{eqn:mat-part}). 
In this way, we can reduce all the induced interactions to those in 
the original potential. 

For a while, we work on a generic potential (\ref{eqn:mat-part}).
The partition function (\ref{eqn:mat-part}) is expressed as an integral
and we can freely reparametrize the integration
variable $\phi$ without changing the partition function.
Obviously only reparametrizations which are analytic at the origin
in the field space are allowed.
If we perform reparametrizations 
$  \phi' = \phi + \epsilon \phi^{n+1} \ (n \geq -1) $
for (\ref{eqn:mat-part}), the coupling constants receive shifts.
This reparametrization invariance implies identities
\bea
  \sum_{i=0}^{n} 
  \left\langle   \frac{1}{N} \tr \phi^i 
  \frac{1}{N} \tr \phi^{n-i} \right\rangle 
  &=&
  \left\langle 
  \frac{1}{N} \tr\left( \phi^{n+1} V'(\phi) \right)
\right\rangle \nonumber \\
&=& \sum^m_{k=1} g_k \la \oon \tr \phi^{n+k}\ra \ \ (n \geq -1).
\label{eqn:sd-eq}
\eea
We note that eq.(\ref{eqn:sd-eq}) constitutes
the first $(K=0)$ set of full reparametrization identities
(discrete Schwinger-Dyson equations)
\cite{FuKaNa:universal,ItMa:virasoro}
\begin{equation}
  \int d \phi \tr \frac{d}{d\phi}  
  \left\{ \phi^{n+1} \prod_{k=1}^K \tr \phi^{n_k}
  \exp \left[ -N\tr V(\phi) \right] \right\} =0, \ \ \ \ (n \geq -1).
  \label{eqn:sd}
\end{equation}
Only reparametrization identities
in the large-$N$ limit will turn out to be necessary in our context.
When we truncate the full set of identities (\ref{eqn:sd})
to those in the large-$N$ limit, 
eq.(\ref{eqn:sd-eq}) exhausts all independent relations,
due to the factorization property (\ref{eqn:fact}).

Eq.(\ref{eqn:sd-eq}) 
can be neatly expressed by introducing the
resolvent operator \cite{BrItPaZu:planar}
\begin{equation}
  \hat{W}(z) = \frac 1N \tr \frac{1}{z \unity - \phi}.
  \label{eqn:def-resolvent}
\end{equation}
The expectation value $\langle \hat{W}(z)\rangle$ is the generating
function of one-point functions:
\begin{equation}
 \langle \hat{W}(z) \rangle =
 \sum_{j=0}^\infty z^{-1-j} \la \frac 1N \tr \phi^j \ra.
\end{equation}
Using the resolvent, eq.(\ref{eqn:sd-eq}) is summarized into a single
equality for $\la \hat{W}(z)\ra$ \cite{Kazakov:kazakovseries},
\begin{eqnarray}
  \lefteqn{\la \hat{W}(z) \ra^2 -V'(z) \la \hat{W}(z)\ra +Q(z)=0,}
  \label{eqn:mat-resolvent-poly}\\
  Q(z)
  & \equiv & 
  \sum_{k=1}^{m-1} \frac{V^{(k+1)}(z)}{k!}
 \left\langle \frac{1}{N} \tr (\phi-z)^{k-1} \right\rangle 
 \nonumber  \\
  & = & 
  \sum_{k=1}^{m-1} \frac{V^{(k+1)}(z)}{k!}
  \sum_{j=0}^{k-1} (-z)^{k-1-j} % correction
  \left( 
  \begin{array}{c}
    k-1\\
    j  \\
  \end{array}
  \right)
  \la \frac{1}{N} \tr\phi^j \ra.
    \label{eqn:mat-q-function}
\end{eqnarray}
Eq.(\ref{eqn:mat-resolvent-poly}) is usually referred to as
 the loop equation.
In the above, $Q(z)$ consists of $\langle {\rm tr} \phi^k \rangle $
for $1 \leq k \leq m-2$.
For our purpose, one-point functions 
$\langle {\rm tr} \phi^k \rangle$ for $3 \leq k \leq m$ are to be regarded
as independent variables, corresponding to the coupling constants
present in the original potential. 
Therefore a special treatment must be applied for 
$\langle {\rm tr} \phi \rangle$ and
$\langle {\rm tr} \phi^2 \rangle$;
we should replace them in terms of $\langle {\rm tr} \phi^k \rangle$
for $3\le k \le m$ with the help of the first two equations of
 (\ref{eqn:sd-eq}) 
\renewcommand{\theequation}{\arabic{section}.\arabic{equation}\alph{subeqn}}
\setcounter{subeqn}{1}
\begin{eqnarray}
  0 & = & \left\langle \frac{1}{N} \tr \phi \right\rangle +
  \sum_{k=3}^{m} g_k \left\langle \frac{1}{N} \tr \phi^{k-1} 
\right\rangle, 
  \label{eqn:mat-first-reparam}  \\
\addtocounter{equation}{-1}
\addtocounter{subeqn}{1}
  1 & = & \left\langle \frac{1}{N} \tr \phi^2 \right\rangle
+\sum_{k=3}^{m} g_k \left\langle \frac{1}{N} \tr  \phi^{k} \right\rangle.
  \label{eqn:mat-second-reparam}
\end{eqnarray}
\renewcommand{\theequation}{\arabic{section}.\arabic{equation}}%
Now let us denote the one-point function of a single trace as 
$ a_j=(1/j) \la (1/N) {\rm tr} \phi^j \ra  $,
and that of the resolvent
with the above replacement as $W$.
It satisfies
\begin{equation}
  W(z;g_j;a_j)^2 -
V'(z)W(z;g_j;a_j)  
+Q(z;g_j;a_j)=0,
\label{eqn:hatw-def}
\end{equation}
with $Q(z;g_j;a_j)$ given by
\begin{eqnarray} % correction
\lefteqn{Q(z;g_3,\ldots,g_m;a_3,\ldots,a_m)} \nonumber \\
& = & \sum_{k=1}^{m-1} V^{(k+1)}(z) (-z)^{k-1} \frac{1}{k!}\nonumber \\
& + & \sum_{k=2}^{m-1} V^{(k+1)}(z) (-z)^{k-1-1} \frac{k-1}{k!}
  (-1) \left(
\sum_{\ell=4}^m (\ell-1) g_\ell a_{\ell-1}  
+  g_3 \left(1-\sum_{\ell=3}^m \ell g_\ell a_\ell \right) 
 \right) \nonumber \\
& + &  \sum_{k=3}^{m-1} V^{(k+1)}(z) (-z)^{k-1-2} \frac{1}{k!}
      \frac{(k-1)(k-2)}{2\cdot 2  }
       \left(1 - 
       \sum_{\ell=3}^m \ell g_\ell a_\ell \right) \nonumber \\
& + &  \sum_{j=3}^{m-2}\sum_{k=j+1}^{m-1} V^{(k+1)}(z) (-z)^{k-1-j} 
\frac{1}{k!}
       \left(
\begin{array}{cc}
k-1\\
j \\
\end{array}
\right)   j a_j .
\label{eqn:def-w-function}
\end{eqnarray}
The quadratic nature of the loop equation (\ref{eqn:hatw-def}) is a 
peculiarity of the one-matrix models, 
as compared to the 
multi-matrix models.
For the case $m=3$, 
the explicit form of $W(z,g,a\equiv a_3)$ reads
\begin{equation}
  W(z,g,a) = 
  \frac 12 \left( z+gz^2- 
     \sqrt{(z+gz^2 )^2   -4 ( 1+gz-g^2+3g^3 a)} 
               \right).
  \label{eqn:cubic-resolvent}
\end{equation}
Note that, by definition (\ref{eqn:def-resolvent}),
only a solution with the large-$z$ asymptotic behavior
$ W \left( z \right) \sim 1/z $ is allowed.
Moreover,  the eigenvalue density distribution $\rho (z)$,
related to $W(z)$ by
$\rho (z)=-1/\pi\ {\rm Im} W(z)$,
should satisfy the trivial identity
\beq
 \int \rho (z) dz =1.
\eeq
We have observed that these two requirements are enough to
determine $W$ without ambiguity for the present case, 
as well as in the case of the two-matrix model.

Another simplification occurs for the averaged form of the saddle 
point equation (\ref{eqn:spe}) 
\begin{equation}
\tas +g \tas^2 -  
W\left(-\frac 1g- \tas ,g, a \right) =0.
  \label{eqn:mat-spe-expect}
\end{equation}
By combining it with
the loop equation (\ref{eqn:hatw-def}), we explicitly obtain $\tas$ 
in the form
\begin{equation}
  \tas = \la \frac 1N  \tr \phi \ra
          = -g + 3 g^2 a \equiv 
 \bar{\alpha}(g,a).
\label{eqn:reduced-spe}
\end{equation}
This is in fact an expected 
result, because 
$\la \as \ra \equiv \langle \Phi^{\rm s}_{N+1,N+1} \rangle
 $ should be equal to
$\la \frac 1N  \tr \phi \ra \equiv \la (1/N)
 \sum_{i=1}^N \Phi_{ii} \ra$.
This fact helps computation in the coset decomposition method considerably, 
for the case of two-matrix model as well.

Using eqs.(\ref{eqn:sp-differential}), 
(\ref{eqn:cubic-resolvent}) and (\ref{eqn:reduced-spe}), 
we obtain a differential equation obeyed by the free energy $F(N,g)$
of the one-matrix model with cubic coupling,
\beq
 \left[ N\frac{\partial}{\partial N} + 2 \right] F(N,g)
 = G\left(g, \frac{\partial F}{\partial g}\right)  + \OO\lt \oon\rt,
 \label{eqn:mat-nrge}  
\eeq
\bea
  G(g,a) & = &
   - \frac{g}{2}  a
   + \frac12 \bar{\alpha}(g,a)^2
   + \frac g3 \bar{\alpha}(g,a)^3
   + \log \lt 1 + g \bar{\alpha}(g,a) \rt \nonumber\\
& +&   \int^{-1/g - \bar{\alpha}(g,a)}_{-\infty} dz
                               \left(W(z,g,a) - \frac{1}{z}\right).
\nonumber \label{eqn:G-function} 
\end{eqnarray}
Since the above equation describes the response under the
shift $N \rightarrow N+1$ it deserves the name of RG equation,
although it is (non-polynomially) nonlinear unlike the form 
anticipated in eq.(\ref{eqn:rge-formal}).\\

\noindent {\large \underline{Comment on the nonlinearity}}

One might wonder if we can keep the equation linear at the cost
of having an infinite dimensional coupling constant space.
We could generalize the partition function to include interactions
such as $\tr \phi^{n_1}\tr\phi^{n_2}\cdots$,
\begin{eqnarray}
  Z_N(g_{n_1 n_2 \cdots}) & = & \int d \phi\ \exp \left[-N S(\phi) \right] ,
\label{eqn:generic-pot}
 \\
  S(\phi) & = & \sum_{k=1}^{\infty} \ \ \sum_{(n_1,n_2,\ldots,n_k)}
  g_{n_1n_2\ldots n_k} N^{1-k} \prod_{i=1}^{k} n_i^{-1} 
  \tr \phi^{n_i}.
  \nonumber
\end{eqnarray}
and enlarge the coupling constant space to
$ 
\{ (g_1,g_2,\ldots, g_{11},g_{12},\ldots,g_{n_1n_2\cdots n_k},\ldots)\}
$.
Then we would expand the right hand side of 
eq.(\ref{eqn:sp-differential})
in terms of 
\begin{equation}
\left.
\left\langle \prod_{i=1}^k \frac 1N \tr \phi^{n_i} \right\rangle
\right|_{(g_j)=(0,1,g,0,0,\ldots)}
= 
\left(\prod_{i=1}^k n_i\right) \cdot
\frac{\partial F}{\partial g_{n_1\cdots n_k}}(N,0,1,g,0,0,\ldots)
\end{equation}
to obtain a linear `RG equation' of the form
\begin{equation}
    \left[ N\frac{\partial}{\partial N} + 2 \right] F(N,0,1,g,0,0,\ldots)
 = \sum_{(n_1,\ldots,n_k)} \beta_{n_1\cdots n_k}(g)
 \frac{\partial F}{\partial g_{n_1\cdots n_k}}(N,0,1,g,0,0,\ldots).
 \label{eqn:mat-trtr}
\end{equation}
It is straightforward to check that there exists no simultaneous zero
for each function $\beta_{n_1\cdots n_k}(g)$.
This contradicts the exact solution of 
the one-matrix model, which has a nontrivial fixed point at
$ g_1=0, g_2=1, g_3=432^{-1/4}, g_j\equiv 0\ (j\ge 4). $
The obvious reason for this apparent contradiction is
that the $(\partial F/\partial g_{n_1\cdots n_k})|_{(g_j)=(0,1,g,0,0,\ldots)}$ 
are not independent quantities in the large-$N$ limit. 
Rather, they are mutually dependent through an infinite number of
relations (\ref{eqn:fact}) and (\ref{eqn:sd-eq}) and are therefore
`$\beta$-functions' for $ g_{n_1\cdots n_k} $.
Accordingly we are forced to take the reparametrization 
identities into account to exclude this ambiguity of $\beta$-functions.

\subsection{One-matrix model II: eigenvalue method}

The RG transformation by means of the 
coset decomposition is in principle applicable 
for an arbitrary potential as well. 
We can eliminate terms such as $(v^\dagger v)^2$,
which are present in the quartic potential case, 
by introducing auxiliary fields.
However, for the model which admits an eigenvalue representation
via $U(N)$-integration we have a more convenient
RG transformation:
integration over the $(N+1)$-th eigenvalue, the extra degree of freedom.
This procedure is particularly useful for one-matrix models with
a generic potential ($ m \ge 4$),
since it can be applied to these models on the same footing.

We can integrate over the $U(N)$ variables in (\ref{eqn:mat-part})
to obtain an integral over 
the eigenvalues $\{\lambda_j\}$ \cite{BrItPaZu:planar}
\begin{equation}
  Z_N(g_j)=c_N \int \prod^N_{i=1} d\lambda_i \ \Delta_N (\lambda_i)^2
  \ \exp \left[-N \sum_{i=1}^N V(\lambda_i) \right],
\end{equation}
where $\Delta_N$ denotes the Van der Monde determinant
$
  \Delta_N(\lambda_j)=\prod_{1\leq i < j\leq N}(\lambda_i-\lambda_j)
$
and  $c_N$ the volume of the $U(N)$ group
$
  c_N=\pi^{N(N-1)/2}/\prod_{p=1}^{N} p!.  
$

In order to relate $Z_{N+1}$ with $Z_{N}$, we shall integrate over  
the $(N+1)$-th eigenvalue $\lambda \equiv \lambda_{N+1}$ in $Z_{N+1}$ 
\begin{eqnarray}
  Z_{N+1}(g_j) & = &  
  \int d\Phi \ \exp \left[-(N+1)\tr V(\Phi) \right] \\
  &=&
  c_{N+1} \int \prod_{i=1}^{N} d\lambda_i \ \Delta_{N}^2 (\lambda_j) 
  \ \e^{ -(N+1) \sum_{i=1}^{N} V(\lambda_i)}
  \int d\lambda  \prod^N_{i=1} \vert \lambda  -\lambda_i \vert^2 
  \ {\rm e}^{-(N+1) V(\lambda )} \nonumber \\
  &=& \frac{c_{N+1}}{c_N} 
  \int d \phi \ \e^{ -(N+1)\tr V(\phi)}
  \int d\lambda 
  \ \exp \left[-(N+1) V(\lambda)+
  2\tr \log \vert \lambda \unity-\phi  \vert \right] \nonumber.
\end{eqnarray}
The $\lambda$-integral can be evaluated by the saddle point 
method as a power series in $1/N$ 
around the saddle point, since the effective potential 
$(N+1)V(\lambda)-2\tr \log | \lambda\unity-\phi |$ is of order $O(N^1)$.
The saddle point equation 
\begin{equation}
  V'(\lambda_{\rm s})
  = \frac 2N \tr \frac{1}{\lambda_{\rm s}\unity -\phi} =
2 \sum_{n=0}^\infty    
  \lambda_{\rm s}^{-n-1}\oon \tr \phi^n  
  \label{6}
\end{equation}
implicitly 
determines the saddle point $\lambda_{\rm s}=\lambda_{\rm s}(g_j,\phi)$
as a $U(N)$-invariant function in $\phi$.
By substituting $\lambda_{\rm s}$ into the partition function, we find 
\bea
  \frac{Z_{N+1}(g_j)}{Z_N(g_j)}&=&
  \frac{c_{N+1}}{c_N}
  \left\langle \exp \left[ -\tr V(\phi)
  -N V(\lambda_{\rm s})+2\tr \log \vert \lambda_{\rm s}\unity-\phi \vert 
  +O(N^0) \right] \right\rangle 
    \label{eqn:recurel} \\
&=&\frac{c_{N+1}}{c_{N}}
  \exp \left[ -\left\langle \tr V(\phi) \right\rangle 
  - N V(\la \lambda_{\rm s}\ra)+2\left\langle 
  \tr\log \vert \la \lambda_{\rm s}\ra \unity -\phi \vert\right\rangle 
  +\OO(N^0) \right].
  \nonumber
\end{eqnarray}
Here the factorization property (\ref{eqn:fact}) is used in the
second line, and 
$\la \lambda_{\rm s} \ra \equiv
\la \lambda_{\rm s}(g_j,\phi) \ra$
is determined again using the resolvent (\ref{eqn:def-resolvent})
\begin{equation}
  V'(\la \lambda_{\rm s} \ra)
  =2\left\langle \frac{1}{N} \tr 
  \frac{1}{\la \lambda_{\rm s} \ra \unity -\phi}\ra
  =2 \la \hat{W}(\la \lambda_{\rm s} \ra)\ra.
  \label{9}
\end{equation}
By taking the large-$N$ limit, 
we get the following differential equation 
as an RG equation  
for the free energy,
\begin{eqnarray}
  \lefteqn{
    \left[ N\frac{\partial}{\partial N} +2 \right] F(N,g_j)}
  \label{11}\\
  & = &  -\frac32 +\left\langle \frac{1}{N} \tr V(\phi) \right\rangle
  +V( \la \lambda_{\rm s} \ra)
  -2\left\langle \frac{1}{N} \tr \log \left\vert \la
  \lambda_{\rm s} \ra \unity
  -\phi \right\vert\right\rangle +O\left(\oon\right).\nonumber
\end{eqnarray}

As in the case of the coset method, 
we can express the right hand side of eq.(\ref{11}) in terms of 
$g_j$ and $a_j=\pa F / \pa g_j$ for $j=3,\cdots,m$
with the help of reparametrization identities.
It is straightforward to express $\tls$ in terms of these quantities,
since the saddle point equation (\ref{9}) readily takes the form
$V'(\tls)=2 W(\tls;g_j;a_j)$.
By combining it 
with the loop equation
(\ref{eqn:hatw-def}), we find an equivalent and
more useful expression
\begin{equation}
  V'(\tls)^2 - 4 Q(\tls;g_j;a_j)=0.
  \label{eqn:red-spe2}
\end{equation}
Eq.(\ref{eqn:red-spe2}) shows that 
$\la \lambda_{\rm s} \ra $ falls on one of the edges of the eigenvalue
distribution $\rho(\lambda)$, because the left hand side is exactly the
descriminant of the loop equation (\ref{eqn:hatw-def}).
This fact can be understood as follows. We adopt the picture
that the $N+1$ eigenvalues, confined by the potential $V(\lambda)$,
interact with each other via a
repulsive Coulomb potential $\log |\lambda_i - \lambda_j|$.
Consider the effective potential for the $(N+1)$-th
eigenvalue, 
which is generated by other $N$ eigenvalues
obeying the distribution function $\rho(\lambda)$.
If the $(N+1)$-th eigenvalue falls on a point $\rho(\lambda)\ne 0$,
it costs an energy loss (of order $N^{-1}\log N$ compared to the total
energy) from the Coulomb interaction.
On the other hand, if the total action is to be minimized when
the $(N+1)$-th eigenvalue is placed on an isolated point outside
the support of $\rho(\lambda)$, the original distribution
is unlikely to minimize the action for $N$ eigenvalues.
Consequently the $(N+1)$-th eigenvalue should fall just on one of the
edges of $\rho(\lambda)$.
Moreover, the true saddle point 
should converge in the $g\rightarrow 0$ limit to $2$ % correction $2\sqrt{2}$
(Wigner distribution), and among such roots it should
minimize the action 
$S(\lambda)\equiv V(\lambda)-
\la (2/N){\rm tr} \log \vert \lambda{\bf 1}-\phi \vert \ra $.
We can always select the true saddle point out of
the $2m-2$ roots of eq.(\ref{eqn:red-spe2}) by imposing
these two requirements, and denote it as 
$\bar{\lambda}=\bar{\lambda}(g_i;a_i)$.

In this way, we can express 
the right hand side of eq.(\ref{11}) in terms of $g_i$ and
$\partial F/\partial g_i$ ($i=3,\ldots,m$),
\begin{eqnarray}
\!\!\!\!\!
  \lefteqn{\left[ N\frac{\partial}{\partial N} +2 \right]
    F(N,g_j)  =   
G_{\rm eig} \left( g_3,\ldots,g_m; 
    \frac{\partial F}{\partial g_3},
    \ldots ,
    \frac{\partial F}{\partial g_m}
  \right) + \OO\lt\oon\rt,} \nonumber \\
\!\!\!\!\!
\lefteqn{G_{\rm eig}(g_3,\ldots,g_m;a_3,\ldots,a_m)}  
\label{eqn:explicit-rge} \\
&\equiv  & 
-1- \sum^m_{k = 3}{k-2 \over 2} g_k a_k
+V(\bar{\lambda}) - 2\log \bar{\lambda} \nonumber\\
& & - 2 \int^{\bar{\lambda}}_{\pm\infty} 
dz \left( W(z;g_3,\ldots,g_m;a_3,\ldots,a_m)-\frac{1}{z} \right).
\nonumber
\end{eqnarray}
The complete set of nonlinear 
RG equations for the one-matrix model consists of
the above equations
(\ref{eqn:explicit-rge}), (\ref{eqn:hatw-def}), and (\ref{eqn:red-spe2}).
So far we have much numerical evidence for the equality of 
the function $G$ in the coset method and the function $G_{\rm eig}$ 
in the eigenvalue method, 
but have not been successful in showing this analytically. 

\subsection{Two-matrix model}

We consider a two-matrix model defined by 
\begin{eqnarray}
  Z_N(g_+,g_-,c) & \!\!\!  = \!\!\! & \int d \phip d \phim
  \exp \left[-N  \tr ( V(\phip,g_+) + V(\phim,g_-) + c\phip \phim)\right],
\nonumber\\
  & & V(x,g)= \frac12 x^2 + \frac g3 x^3,
  \label{eqn:2mm-cubic-pot}
\end{eqnarray}
where $\phipm$ are $N\times N$ hermitian matrices.
The free energy 
\begin{equation}
  F(N,g_+,g_-,c) = -\frac{1}{N^2} \log
  \frac{Z_N(g_+,g_-,c)}{Z_N(0,0,0)}
\label{eqn:freeenergytm}
\end{equation}
represents 
the partition function of the Ising model on random 
triangulated surfaces \cite{Kazakov:Ising},
and the Ising temperature $T$ is proportional to $-\log^{-1} (-c)$.
This model is also known to allow an eigenvalue representation
\cite{ItZu:2mm}
\bea
Z_N(g_+,g_-,c)& \!\!\!  = \!\!\! &
\frac{(-c\pi^2)^{N(N-1)/2}}{\prod _1^N p!}
\int \prod_{i=1}^N d \lambda_{+,i}  d \lambda_{-,i}\ 
\Delta_N(\lambda_+) \Delta_N(\lambda_-)\nonumber \\
& \!\!\! \cdot \!\!\! &
\exp \left[-N  \sum_{j=1}^N 
( V(\lambda_{+,j},g_+) + V(\lambda_{-,j},g_-) + 
c \lambda_{+,j} \lambda_{-,j}  ) \right] .
\eea
However, 
the integrand is an alternating function in $\lambda_{\pm,i}$,
since the Van der Monde determinants
is not squared but bi-linear in the above expression.
Therefore it is inappropriate
to evaluate the $\lambda_{\pm,N+1}$-integral by the saddle point method.
This argument also applies to multi-matrix models with a chain-like 
interaction
$c \sum_{a=1}^{p-1} \tr \phi_a \phi_{a+1}$.
For this reason we present only the coset method in the following.\\

\noindent {\large \underline{RG transformation}}

We decompose $(N+1)\times (N+1)$ hermitian matrices 
$\Phi_\pm$ into
\begin{equation}
    \Phi_{\pm}= \left(
  \begin{array}{@{\,}c|c@{\,}}
  \PHIPM         & \VPM         \\ \hline
   v_\pm^\dagger & \alpha_\pm   \\
  \end{array}
  \right)
\end{equation} 
as with the case of one-matrix models.
In terms of these variables, the partition function
(\ref{eqn:2mm-cubic-pot}) reads
\begin{eqnarray}
  \lefteqn{ Z_{N+1} (g_+,g_-,c)} \nonumber \\ 
  \hspace*{-5em}
  & = & \!\!\!
 \int d\phip d\phim dv_+ dv_+^\dagger dv_- dv_-^\dagger  d\alpha_+ d\alpha_-
\label{eqn:2mm-decomp-part} \\
  & &  
  \cdot \exp [-(N+1)\tr(V(\phip,g_+) + V(\phim,g_-) +  c \phip \phim) ] 
\nonumber \\
  & &   
  \cdot \exp [-(N+1)(V(\alpha_+,g_+) +V (\alpha_-,g_-)+c \alpha_+ 
\alpha_-)]
\nonumber \\
  & &   
  \cdot \exp \left[-(N+1)\cdot
  (  v_+^\dagger  v_-^\dagger)
  \left(
  \begin{array}{cc}
    \unity +g_+(\phip + \alpha_+ \unity) & c \unity \\
    c \unity       &       \unity +g_-(\phim + \alpha_- \unity) \\
  \end{array}
\right)
\left(
\begin{array}{l}
  v_+ \\
  v_- \\
\end{array} 
\right)
\right].
\nonumber
\end{eqnarray}
The part of the action quadratic in  $(v_+,v_-)$ and
$(v_+^\dagger,v_-^\dagger)$ 
is presented in the form of a $2N\times 2N$ matrix. 
We perform the integration over
$(v_+,v_-)$-field to obtain
\begin{eqnarray}
  Z_{N+1}(g_+,g_-,c) & = &
  \left(\frac{\pi}{N+1}\right)^{2N}
  \int d\phip d\phim  \\
  & & \cdot \exp \left[-(N+1)\tr(V(\phip,g_+)+V(\phim,g_-) +  c\phip \phim )
\right]  
  \nonumber \\ 
  & & \cdot \int d\alpha_+ d\alpha_-
  \exp[-(N+1) (V(\alpha_+,g_+) + V(\alpha_-,g_-) + c \alpha_+ \alpha_-)]
 \nonumber \\
  & &   \cdot \exp\left[
  -\Tr\log \left(
  \begin{array}{cc}
    \unity +g_+(\phip + \alpha_+ \unity) & c \unity \\
    c \unity       &       \unity +g_-(\phim + \alpha_- \unity) \\
  \end{array}
\right)\right]    \nonumber.
\end{eqnarray}
The trace $\Tr$ is taken over a $2N \times 2N$  matrix, whereas $\tr$ 
is over an $N \times N$ matrix.
Again we employ the saddle point method for
the $\alpha_\pm$-integration.
The saddle point $\asa{\pm}=\asa{\pm}(g_+,g_-,\phip,\phim)$ 
is determined by the equations
\begin{equation}
  \asa{\pm} + c\ \asa{\mp} + g_\pm {\asa{\pm}}^2 + 
  g_\pm  \frac{1}{N} \Tr_\pm 
  \left(
  \begin{array}{cc}
    \unity +g_+(\phip + \asa{+} \unity) & c \unity \\
    c \unity       &       \unity +g_-(\phim + \asa{-} \unity) \\
  \end{array}
\right)^{-1}=0,
\label{eqn:saddle-alphapm}
\end{equation}
where
\begin{equation}
   \Tr_\pm M \equiv \Tr ( P_\pm M ),\ \ \ 
   P_+=
   \left(
  \begin{array}{cc}
    \unity & 0 \\
    0 & 0 \\
  \end{array}
  \right) ,\ \ 
  P_-=
   \left(
  \begin{array}{cc}
    0 & 0 \\
    0 & \unity \\
  \end{array}
  \right) .
\end{equation}

As in the case of the one-matrix model, we employ the factorization
property (\ref{eqn:fact}) and approximate 
$Z_{N+1}/Z_N$ by a differential to obtain
\begin{eqnarray}
& & \left[ N\frac{\partial}{\partial N} + 2 \right] F(N,g_+,g_-,c) 
\label{242}\\
& & =  - 1 +  \la \frac 1N\tr V(\phip,g_+)\ra + 
              \la \frac 1N\tr V(\phim,g_-)\ra +
           c\ \la \frac 1N\tr \p_+ \p_-   \ra \nonumber \\ 
& & + 
  V(\la \asa{+} \ra ,g_+) + V(\la \asa{-} \ra ,g_-) +
  c \la \asa{+} \ra \la \asa{-} \ra \nonumber \\
& & +   \la \frac 1N\Tr \log 
  \left(
    \begin{array}{cc}
     \unity +g_+(\phip + \la \asa{+} \ra  \unity)   & c \unity   \\
     c \unity    &    \unity +g_-(\phim + \la \asa{-} \ra \unity) 
  \end{array}
  \right)
\ra + \OO\lt \oon \rt ,\nonumber
\end{eqnarray}
where
\begin{equation}
  \langle \cdots \rangle = Z_N^{-1} 
  \int d\phip d\phim (\cdots) \exp [-N 
  \tr (V(\phip,g_+)+V(\phim,g_-)+c\phip\phim) ].
\end{equation}

\noindent {\large \underline{Reparametrization invariance}}

In what follows we concentrate on the
case $g_+=g_-\equiv g$ which describes 
the Ising model coupled to two-dimensional quantum gravity in
the absence of an external magnetic field.
In addition, we assume that 
\begin{equation}
  \langle \tr \phip^i \rangle = \langle \tr \phim^i \rangle
  \label{eqn:symmetric-1pt-fn}
\end{equation}
holds for any $i$.
When we trace on a part ($T\leq T_{\rm cr}$) of the critical line, 
this assumption amounts to defining
the free energy as the sum of contributions from 
two spontaneously magnetized states.
A direct consequence of this assumption is
\begin{equation}
  \la \alpha_{\rm s,+} \ra = \la \alpha_{\rm s,-} \ra  
\equiv  \la \alpha_{\rm s} \ra .
  \label{eqn:symmetric-alpha}
\end{equation}
Under these conditions, the saddle point equation for $\tas$
reduces to the form
\begin{equation}
  \tas + c \tas + g \tas^2 + 
  \left\langle \frac{1}{2N}
  \Tr 
  \frac{1}{ (1/g +  \tas )\bs_0 + (c/g)   \bs_1  + \bp} 
\right\rangle = 0,
\label{eqn:sadpttwomatrix}
\end{equation}
where 
\begin{equation}
  \bp =  \left(
  \begin{array}{cc}
    \phip & 0 \\
    0     & \phim \\
  \end{array}
\right), \ \ 
  \bs_0= \left(
  \begin{array}{cc}
    \unity & 0 \\
    0      & \unity \\
  \end{array} \right), \ \ 
  \bs_1= \left(
  \begin{array}{cc}
    0 & \unity  \\
    \unity & 0 \\
  \end{array} \right).
\end{equation}
The differential equation (\ref{242}) simplifies into
\begin{eqnarray}
  \lefteqn{\left[ N\frac{\partial}{\partial N} + 2 \right]  F(N,g,c)}
  \nonumber \\
  & = & - 1 + \la \frac 1N  \tr V(\phip,g) \ra +
              \la \frac 1N  \tr V(\phim,g) \ra +
  c\la \oon \tr \phip\phim\ra 
\label{eqn:rengrtwomatrix} \\
 &+& 2 V(\tas,g) +c \la \as \ra^2 + \left\langle \frac1N \Tr \log 
  \lt \lt \frac1g + \tas \rt \bs_0 + \frac cg \bs_1 + \bp \rt
\right\rangle + \OO\lt \oon \rt.
\nonumber
\end{eqnarray}

The right hand side of eq.(\ref{eqn:rengrtwomatrix}) 
consists of terms of the form
$\tr (\phip^{n_1} \phim^{m_1} \phip^{n_2} \phim^{m_2} \cdots)$.
We observe 
that all these induced interactions
can be reduced solely to the interactions present in the original
potential, i.~e.\ $\tr (\phip^3+\phim^3)$ and $\tr \phip \phim$,
by recursively using the reparametrization identities
\bea
\int d \phip d \phim &\!\!\!\!\!\! & \tr \frac{d}{d\phi_\pm} \left\{
 \phip^{n_1} \phim^{m_1} \phip^{n_2} \phim^{m_2} \cdots \right.
\nonumber \\
&\!\!\! \cdot \!\!\!& \left. 
\exp\left[ -N \tr \lt V(\phip,g)+V(\phim,g)+c \phip \phim \rt \right]
\right\}=0 .
\label{eqn:sdtwomat}
\eea
Again this procedure can be performed
at once by introducing the resolvent operator;
in fact we can find a closed set of loop equations (A.2)
that expresses the one-point function of the
resolvent of the required form (see the appendix A)
\begin{equation}
W_0 (z;g,c; a_g,a_c) =  \left\langle 
\frac 1N \Tr \frac{1}{ z \bs_0 + (c/g) \bs_1  + \bp } \right\rangle 
\label{eqn:resolventtwomm}
\end{equation}
in terms of $a_g=\la (1/3N) \tr (\phip^3+\phim^3) \ra$ and
$a_c=\la (1/N)\tr \phip \phim \ra$.

Furthermore, by combining the saddle point equation 
(\ref{eqn:sadpttwomatrix}), or
\begin{equation}
  (1+c) \tas + g \tas^2 + \frac 12 
W_0 \left(\frac 1g + \tas ;g,c; a_g,a_c \rt
   = 0
\end{equation}
and the loop equation (\ref{2mmloopquart}) we can prove
\beq
\tas = \la \oon \tr \phi_\pm \ra=
\frac{g}{1+c} \lt -1+c\ a_c+\frac{3g}{2} a_g \rt
\equiv 
\bar{\alpha}(g,c;a_g,a_c)
\label{2mm-spe-red} 
\eeq
as expected. These three equations (\ref{eqn:rengrtwomatrix}),
(\ref{2mmloopquart}) and (\ref{2mm-spe-red}) constitute
the nonlinear RG equation for the two-matrix model
\begin{equation}
\left[ N \frac{\partial}{\partial N} + 2 \right]F  
 = G\left(g,c; 
     \frac{\partial F}{\partial g},
     \frac{\partial F}{\partial c}\right)
     + \OO\lt\oon\rt ,
\label{eqn:rgetwomm}
\end{equation}
\begin{eqnarray*}
 G(g,c;a_g,a_c) 
&\!\! \equiv \!\!  &
-  \frac{g}{2} a_g
+ ( 1 + c) \bar{\alpha}^2
         + \frac 23 g \bar{\alpha}^3
+ 2 \log ( 1 + g \bar{\alpha}) \\
& &   + \int_{- \infty}^{1/g+ \bar{\alpha}} dz \ 
     \left( W_0(z;g,c;a_g,a_c) - \frac 2z \right). 
\label{eqn:2mm-G-function} 
\end{eqnarray*}

%We note that the assumption (\ref{eqn:symmetric-1pt-fn})
%is appropriate for the unmagnetized phase.
%On the other hand, for a region of coupling space where spins are
%spontaneously magnetized, this assumption amounts to defining
%the free energy by
%\begin{equation}
%  F(N,g) = - \frac{1}{N^2}\log ( 
%  {\rm e}^{-N^2 F_+} + {\rm e}^{-N^2 F_-}).
%\end{equation}
%Here $F_+$ ($F_-$) denotes the contribution from a saddle point
%corresponding to the upward (downward) magnetized phase.
%If one wants to study the RG equation obeyed by
%the pure state free energy $F_\pm (N,g)$,
%one has to solve the saddle point equation
%(\ref{eqn:saddle-alphapm}) without 
%assuming $g_+=g_-$ or (\ref{eqn:symmetric-1pt-fn}).
%Then we should select an appropriate branch of the 
%loop equation and take the limit 
%$g_+ \rightarrow g_- \pm 0$.

%%%%%% Fixed Pts & Operator Contents %%%%%%%%
\section{Fixed points and operator contents}
\setcounter{equation}{0}
\subsection{Solution to the nonlinear RG equation}
\label{sec:nonlin-algorithm}
In the previous section we have derived nonlinear differential equations,
governing the critical behavior of the free energy $F(N,g)$,
of the form
\begin{equation}
  \left[ N \frac{\partial}{\partial N}  + 2  \right] F(N,g)
  = G\left( g, \frac{\partial F(N,g)}{\partial    g}\right) 
    + \OO\lt\oon\rt
  \label{eqn:nrge}
\end{equation}
(or its multi-variable extension)
instead of the linear one (\ref{eqn:rge-formal}).
To complete the program we now need a formula which relates 
positions and exponents of critical points
to $G$, just as in the linear case eq.(\ref{eqn:mat-exponents}).\\

\noindent {\large \underline{Sphere limit}}

First we concentrate on the leading part $F^0(g)$ of the free
energy in the $1/N$-expansion.  
It is easy to see that $F^0$ satisfies
\begin{equation}
  2 F^0(g) = G\left(g, \frac{\partial F^0(g)}{\partial g}\right)
  \label{eqn:leadingrge}
\end{equation}
We assume that $F^0$ consists of analytic and non-analytic parts 
around a critical point $g_*$,
\begin{equation}
  F^0(g) = \sum_{k=0}^\infty a_k(g-g_*)^k +
           \sum_{k=0}^\infty b_k(g-g_*)^{k+2-\gamma_0} \ \ \ 
(\gamma_0\not\in\Z).
\end{equation}
Quantities $a_k,b_k, \gamma_0, g_*$ are unknown 
and are to be determined. 
In anticipation of the result, we have used the same notation as 
the susceptibility exponent $\gamma_0$ in 
eqs.(\ref{eqn:exact-sol}) and (\ref{eqn:exponenth}). 
We expand the function $G(g,a)$ around 
$(g_*, a_1)$ as follows
\begin{eqnarray}
G(g,a) & = & \sum_{n=0}^\infty \beta_n(g)(a-a_1)^n \nonumber \\
       & = & \sum_{n=0}^\infty \sum_{k=0}^\infty 
\beta_{nk}\cdot (g-g_*)^k(a-a_1)^n.
\label{betan}
\end{eqnarray}
We substitute these expressions into the nonlinear RG equation
(\ref{eqn:leadingrge}) and compare the coefficients of
various powers of $g-g_*$ on both sides.
Since $b_0 \ne 0$ by definition, the most singular term
$(g-g_*)^{1-\gamma_0}$ determines the fixed point 
\renewcommand{\theequation}{\arabic{section}.\arabic{equation}\alph{subeqn}}
\setcounter{subeqn}{1}
\begin{equation}
   0 = \beta_{10} (2-\gamma_0) b_0. \label{eqn:fixedpoint}
\end{equation}
\addtocounter{equation}{-1}%
\addtocounter{subeqn}{1}%
The next singular term $(g-g_*)^{2-\gamma_0}$ determines the critical
exponent $\gamma_0$
\begin{equation}
  2 b_0 = (2-\gamma_0) b_0 ( \beta_{11}  + \beta_{20} \cdot 4 a_2).
  \label{eqn:exponent}
\end{equation}
\addtocounter{equation}{-1}%
\addtocounter{subeqn}{1}%
So far there appear two unknown coefficients $a_1$ and $a_2$ 
from the analytic part. 
However, by comparing terms $(g-g_*)$ and $(g-g_*)^2$, we
can fix these quantities
\begin{eqnarray}
  2 a_1 & = & \beta_{01}, \label{eqn:nonlin-1-reg1} 
\label{eqn:analyticcoeff}\\
\addtocounter{equation}{-1}
\addtocounter{subeqn}{1}
2 a_2 & = & \beta_{02} + \beta_{11} 2 a_2 + \beta_{20}  
(2a_2)^2
                          \label{eqn:nonlin-1-reg2}. 
\end{eqnarray}
These four equations can be rewritten in terms of the function $G(g,a)$:
\setcounter{subeqn}{1}
\begin{eqnarray}
0 & = & G_{,a}(g_* ,a_1),
\label{3.7a}
\\
\addtocounter{equation}{-1}
\addtocounter{subeqn}{1}
\frac{2}{2-\gamma_0} & = & G_{,g,a}(g_* ,a_1)+2a_2
G_{,a,a} (g_*, a_1),
\label{3.7b}
\\
\addtocounter{equation}{-1}
\addtocounter{subeqn}{1}
2 a_1 & = & G_{,g}(g_*, a_1),
\label{3.7c}
\\
\addtocounter{equation}{-1}
\addtocounter{subeqn}{1}
2 a_2 & = & \frac{1}{2} G_{,g,g}(g_*, a_1)+2a_2  G_{,g,a}(g_*, 
a_1)
+ 2 (a_2)^2 \ G_{,a,a} (g_*, a_1) .
\label{3.7d}
\end{eqnarray}
\renewcommand{\theequation}{\arabic{section}.\arabic{equation}}%
Obviously eqs.(\ref{3.7a}) and (\ref{3.7b}) are 
the nonlinear counterparts of the usual equations 
determining
fixed points and scaling exponents, 
\beq
0=\beta(g_*),\ \ {\rm and}\ \ 2/(2-\gamma_0)= \beta'(g_*) ,
\eeq
respectively.
We can solve the above four equations simultaneously
to fix four unknowns,
the critical coupling $g_*$, critical exponent $\gamma_0$, and
coefficients $a_1,a_2$.
Moreover, the other consistency conditions determine all 
the other coefficients at each fixed point,
$a_k$ $(k\ge 0)$ and $b_k/b_0$ ($k\ge 1$) recursively except for the 
overall normalization of the singular term $b_0$. 
Although the equation for $a_2$ is quadratic,
we can choose the correct branch, that which continues to the unique 
solution at the origin (the Gaussian fixed point).
In this way we can obtain the series expansion of the sphere free energy
around the fixed point, up to one integration constant.

It is straightforward to generalize our algorithm to the case of
multi-coupling constants. 
We consider a differential equation
\begin{equation}
\left[ N \frac{\partial}{\partial N} +  2 \right]
F(N,g_j) 
= G\left(g_j;  \frac{\partial F}{\partial g_j} \right)
 + \OO\lt\oon\rt .
\label{eqn:nonlin-m-diff}
\end{equation}
We assume that the leading part  $F^0(N,g_j)$ of the
$1/N$-expansion of a solution
can be expanded in the form ($\delta g_j \equiv g_j - g_{j*}$)
\begin{equation}
F^0(N,g_j) =
\sum^{\infty}_{n=0} a_{j_1 \cdots j_n} 
\delta g_{j_1} \cdots \delta g_{j_n} + 
\sum_i \left( V_{ij} \delta g_j \right)^{2-\gamma_0^{(i)}}
\sum^{\infty}_{n=0}  
b_{(i) j_1 \cdots j_n} \delta g_{j_1} \cdots \delta g_{j_n},
\label{f0rgw}
\end{equation}
where the repeated indices are summed over.
A transformation matrix $V_{ij} $ is introduced to
diagonalize the scaling matrix. 
Then we have a set of equations to determine 
$g_{j*}$, $\gamma_0^{(i)}$, $a_i$ and $a_{ij}$
\renewcommand{\theequation}{\arabic{section}.\arabic{equation}\alph{subeqn}}
\setcounter{subeqn}{1}
\begin{eqnarray}
& & 0  =  \frac{\partial G}{\partial a_i} , \label{eqn:mat-m-fp}
\label{3.10a} \\
\addtocounter{equation}{-1}
\addtocounter{subeqn}{1}
& &\sum_k (V^{-1})_{ik} \frac{2}{2-\gamma_0^{(k)}} V_{kj}
 =  \frac{\partial^2 G}{\partial g_j \partial a_i} 
+2 a_{jk} \frac{\partial^2 G}{\partial a_i \partial a_k}
 \equiv G_{ij},
\label{eqn:mat-m-exp} \\ 
\addtocounter{equation}{-1}
\addtocounter{subeqn}{1}
& &2 a_i  =  \frac{\partial G}{\partial g_i} \\
\addtocounter{equation}{-1}
\addtocounter{subeqn}{1}
& &2 a_{ij}  =  \frac 12 \frac{\partial^2 G}{\partial g_i \partial 
g_j} 
+a_{ik} \frac{\partial^2 G}{\partial g_j \partial a_k} 
+a_{jk} \frac{\partial^2 G}{\partial g_i \partial a_k} 
+2 a_{ik} a_{j\ell} \frac{\partial^2 G}{\partial a_k \partial a_\ell} ,
\label{eqn:mat-m-reg2} \label{eqn:2parameters}
\end{eqnarray}
\renewcommand{\theequation}{\arabic{section}.\arabic{equation}}%
where they stem from terms of order 
$O \left( (V_{ij} \delta g_j)^{1-\gamma_0^{(i)}} \right)$,
$O \left( (V_{ik} \delta g_k)^{1-\gamma_0^{(i)}} (V_{jk} \delta g_k ) 
\right)$,
$O(\delta g_i)$ and $O(\delta g_i \delta g_j)$, 
respectively. 
We refer to the right hand side $G_{ij}$ 
of eq.(\ref{eqn:mat-m-exp}) 
as the `scaling exponent matrix'.\\

\noindent {\large \underline{Exceptional cases}}

So far we have implicitly assumed that 
the exponent $\gamma_0$ is irrational. 
In most cases this assumption is incorrect and
higher powers of $F_{\rm sing}$
appearing in the right hand side of eq.(\ref{eqn:leadingrge})
contribute to the analytic part.
Therefore eqs.(\ref{3.7c}--d)
which serve to determine the coefficients $a_1$ and $a_2$
may require modification on such occasions.
However, for $c<1$ theories that we examine in this paper,
$\gamma_0$ is known to be negative.
Then we observe that the potentially dangerous terms
\beq
\lt \pa F_{\rm sing}/\pa g\rt^k \sim
 O(\delta g^{(1-\gamma_0)k})
\ \ (k\geq 2)
\eeq
are always of higher order than $O(\delta g^2)$.
Hence the mixing of such terms does not occur
in eqs.(\ref{3.7c}--d)
determining $a_1$ and $a_2$ which are
needed to identify fixed points and scaling exponents.

The function $\beta_n(g)$ in eq.(\ref{betan}) 
may sometimes develops singularity.
In such occasions the $G$-function can be separated into
its regular and singular parts around a specific point
according to the power expansion of each 
$\beta_n(g)$.
Then we can prove that our algorithm, which originally assumes 
the analyticity of the $G$-function,
is still valid after a replacement
of $G$ in eqs.(\ref{3.7a}--d) or (\ref{3.10a}--d) with 
its regular part $G_{\rm reg}$.\\

\noindent {\large \underline{Higher genus contributions}}

Now we show 
that eq.(\ref{eqn:nrge}) is sufficient to determine
the singular behavior of higher genus free energies $F^h$.
Suppose we retain all terms of subleading order in $1/N$
in each procedure,
i.~e.~saddle point evaluation of the $\alpha$- or 
$\lambda$-integral,
application of the reparametrization identities, and
replacement of a difference in $N$ by differentials.
Then we can express this complete RG equation in the form
\beq
 \left[ N\frac{\partial}{\partial N} + 2 \right] F(N,g)
 = G\left(g, \frac{\partial F}{\partial g}\right)  + 
 \Delta G(N,g;[F]),
 \label{completeRGeq}  
\eeq
where $[F]$ denotes differential polynomials of $F(N,g)$.

The partial differential equation (\ref{completeRGeq}) can be
separated into a set of ordinary 
differential equations for each genus contribution. 
It is important to realize that the additional contribution
$\Delta G$
introduced into the right hand side of the RG 
equation carry additional powers of $1/N$. 
Therefore they will not appear in the coefficient $\beta_n(g)$ 
\begin{equation}
   ( 2 - 2 h) F^h(g) =   r^h(g)
  + \sum_{n=1}^{\infty} 
  \beta_n(g) \sum_{\stackrel{\scriptstyle h_1+\cdots+h_n=h,}{h_i\geq 0}}
  \prod_{i=1}^{n} \frac{d F^{h_i}}{dg}.
\label{exp}
\end{equation}
The inhomogeneous term $r^h(g)$ does get various higher order 
contributions including those terms from $F^{h'}$ for $h' \le h-1$. 

Now we assume that
$F^{h'} \ \ (h' \le h-1)$ is less singular
than $F^h$. 
Then the leading singular part in the right hand side of 
eq.(\ref{exp}) is dominated by the terms involving
$dF^h/dg$.
Once we realize this structure of the RG equation, 
we can repeat the same argument as in the case of the sphere to obtain 
the scaling behavior of the higher genus contributions. 
The singularity of $F^{h'} \ \ (h' \le h-1)$ does not affect 
eqs.(\ref{eqn:fixedpoint}--d).
We find that the fixed point condition is the same as 
eq.(\ref{eqn:fixedpoint}) except that the coefficient $b_0$ is now 
replaced by $b_0^h$ of the singular term of the genus $h$ free energy. 
The critical exponent condition becomes 
\begin{equation}
2(1-h)b_0^h  =  b_0^h (2-\Gamma_h) (
\beta_{11} 
+ \beta_{20}\cdot 4a_2  ) .
\label{eqn:hcriticalexponent}
\end{equation}
The coefficients $a_1$ and $a_2$ of 
the analytic term are needed to fix these equations. 
However, it is notable that they are precisely the 
analytic terms in the {\it sphere} free energy. 
Therefore the same conditions as the sphere case 
(\ref{3.7c}) and (\ref{3.7d})
are sufficient to determine them. 
In this way we find that the fixed point is universal for any genus,
and the critical exponent $\Gamma_h$ for genus $h$ is given by 
\begin{equation}
2-\Gamma_h=(1-h)\gamma_1, \qquad \gamma_1+\gamma_0=2.
\label{3.13} 
\end{equation}
This result explains the double scaling behavior for the singular 
part of the free energy 
\begin{eqnarray}
  F_{\rm sing}(N, g) & = &
  \sum_{h=0}^{\infty} N^{-2h}F_{\rm sing}^h(g)  \nonumber \\
  & = & (g - g_{*})^{2- \gamma_0} 
  f\left(N^{2/\gamma_1}(g-g_{*})\right) + \mbox{less singular terms}.
\end{eqnarray} 
Similarly we observe
that the generalization of eq.(\ref{3.13}) to the multi-coupling
case,
\begin{equation}
2-\Gamma^{(k)}_h=(1-h)\gamma^{(k)}_1 ,
\end{equation}
holds for each $k$.

\subsection{Spectrum of continuum theories}
In this subsection we recall the spectrum
of operators of continuum theories of quantum gravity
and present the various scaling exponents,
%and gravitational scaling dimensions
to prepare for the following subsections.

Operators of $(p,q)$-minimal conformal matter
coupled to two-dimensional quantum gravity
(hereafter referred to as the $(p,q)$ gravity)
derived from matrix models can be identified with hamiltonians of
$W_{p}$-constrained, $p$-reduced KP hierarchy 
with a source insertion at $t_{p+q}$ 
\cite{GGPZ:Action,FuKaNa:universal}.
Namely the $k$-th order KP hamiltonian 
${\cal O}_k$ for $k<p+q$ can be identified
with the gravitationally dressed
$(n,m)$-th primary operator with $k=\vert pm-qn \vert$,
\beq
{\cal O}_{\vert pm-qn \vert}
 \sim \int d^2 z\ \Psi_{n,m} \E^{\beta_{n,m} \varphi}
\label{eqn:vertexop}
\eeq
where $\varphi$ denotes the Liouville field.
Operators ${\cal O}_k$ for $k>p+q$ are 
the gravitational descendants.
The $W_p$ constraint implies that the free energy
\beq
F(\kappa;t_i)= \sum_{h=0}^{\infty} \kappa^h
\la \exp {\sum_{k\geq 1} t_k {\cal O}_k} 
\ra_{h,\ \rm conn.}
\eeq
has the following scaling property:
\bea
& &\la \exp \lt t_k {\cal O}_k+ {\rm const.}{\cal O}_{p+q} \rt
\ra_{h,\ \rm conn.}
\sim t_k^{2-\Gamma^{(k)}_h} \ \ \ \ \ (k<p+q), \nonumber\\
& &\Gamma^{(k)}_h = \gamma^{(k)}_0 + h \gamma^{(k)}_1,
\ \ \gamma^{(k)}_0=-\frac{2k}{p+q-k},\ \ 
\gamma^{(k)}_1=\frac{2(p+q)}{p+q-k}.
\label{eqn:kthexp}
\eea
The exponent $\gamma_0^{(k)}$ is related to the more
familiar `gravitational scaling dimension' 
\cite{KnPoZa:LightCone} 
$\Delta^{(k)}$
by
\begin{equation}
  \Delta^{(k)} = 1 - \frac{2-\gamma^{(1)}_0}{2-\gamma^{(k)}_0}.
\end{equation}
These exponents can as well be derived from the Liouville
theory; they are related with the background charge $Q$ and 
the Liouville momentum $\beta_{n,m}$ of a vertex operator 
by 
\bea
& &\gamma^{(\vert pm-qn \vert)}_0=2+\frac{Q}{\beta_{n,m}},\ \ 
\gamma^{(\vert pm-qn \vert)}_1=-\frac{Q}{\beta_{n,m}}, \\
& & Q\equiv\sqrt{\frac{25-c}{3}}=\sqrt{\frac{2}{pq}} (p+q),\nonumber \\ 
& & \beta_{n,m}\equiv -
\frac{\sqrt{25-c}-\sqrt{1-c+24 \Delta_0^{(n,m)}}}{2\sqrt{3}}=
-\frac{p+q-\vert pm-qn \vert}{\sqrt{2pq}}.\nonumber
\eea 
The exponent $\gamma_0^{(1)}$ for the dressed lowest dimensional
operator ${\cal O}_1$ is usually
referred to as the string susceptibility exponent $\Gamma_{\rm str}$.

It is instructive to point out that
the so-called boundary operators \cite{MaMoSe:boundaryops},
${\cal O}_{k}$ for $k=0$ mod $p$ or $q$, 
should be
excluded from the spectrum in our formalism for the following reason.
Consider the $(p,q)$ gravity as the
$W_{p}$-constrained, $p$-reduced KP hierarchy 
with a source insertion at $t_{p+q}$.
Due to the $p$-reduction condition, KP operators
${\cal O}_{k}$ for $k=0\ \mod\ p$ decouple
\beq
\langle {\cal O}_{np} \cdots \rangle \equiv 0.
\eeq
On the other hand, KP operators ${\cal O}_{k}$ 
for $k=0\ \mod\ q$
do not decouple automatically. In turn they can always be absorbed 
into the redefinition of the operators of lower order than them.
For instance, in the case of a unitary theory $(q=p+1)$, 
shifts in the KP time parameters $t_k$ of the form
\bea
t_1 &\too& t'_1=t_1+
\lt \mbox{polynomial in $t_2,\ t_3,\ \cdots,\ t_{p-1},\ t_{p+1}$} \rt
,\nonumber \\
t_2 &\too& t'_2=t_2+
\lt \mbox{polynomial in $t_3,\ \cdots,\ t_{p-1},\ t_{p+1}$} \rt
,\nonumber \\
  & & \cdots    \\
t_{p-1} &\too& t'_{p-1}=t_{p-1}+\lt \mbox{polynomial in $t_{p+1}$} \rt,
\nonumber \\
t_{p+1} &\too& t'_{p+1}=0 \nonumber
\eea
are enough to eliminate $t_{p+1}$ \cite{FuKaNa:pqduality}. 
In this sense these operators are redundant.
This type of shift of parameters (of the continuum theory)
should automatically be dealt with
in our procedure to reduce the coupling space, 
since we have utilized {\it all} possible reparametrization 
invariance at the discrete level.
Consequently KP operators 
${\cal O}_{k}$ for $k=0\ \mod\ p\ {\rm or}\ q$ would never 
appear in the spectrum of our RG equation.
In the subsequent
subsections we will observe that this is indeed the case.
The fact that $p$ and $q$ are treated on equal footing in our 
formalism
should be contrasted with the description by the KP flow.
This feature will also be discussed in sect.4.2.

\subsection{Fixed points of the one-matrix model}
Here we present a full account of the result for one-matrix models, 
part of which has been reported in ref.\cite{HiItNiSa:Matrix}. 
We investigate the case of two couplings ($g_3$ and $g_4$) as an example,
by employing the eigenvalue method.
The exact solution tells us that 
there exists a pure gravity critical line 
parameterized by
\begin{eqnarray}
g_3 & = & \frac{2(32z^2 -z^4-64z w +  4z^3 w+64 w^2)}{z^5+2z^4 w-4z^3 w^2},\ 
g_4  =  \frac{16(16z -z^3-32 w)}{3(z^5+2z^4 w-4z^3 w^2)},
\nonumber \\
0 & = &  -96 z^4 + 3z^6 -6z^5 w - 6z^4 w^2 + 256 z^4 w^3 +
 8z^3w^3-128w^4 
\end{eqnarray}
and the tricritical point at one end of the line. 
(See Fig.1).

\noindent
\begin{minipage}{\textwidth}
\begin{center}
    \leavevmode
    \figinclude{200}{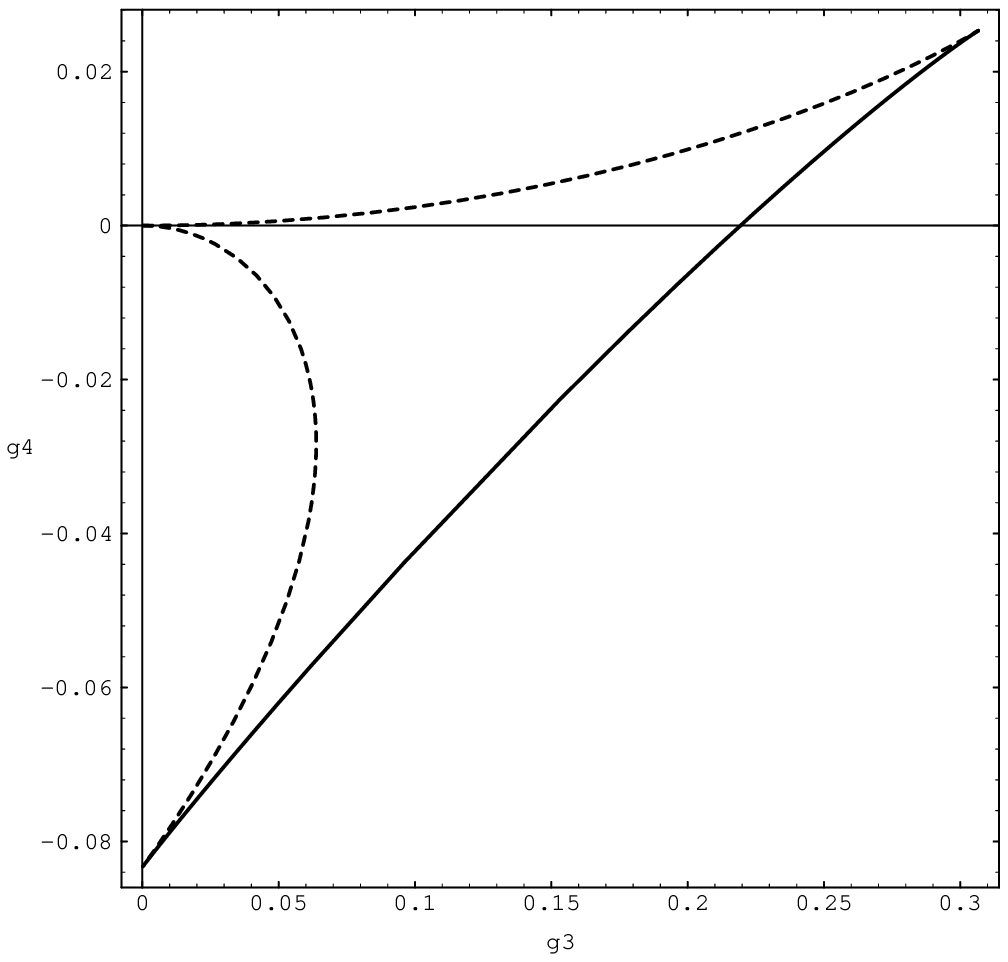}\\
    
{\small Figure 1:
    The pure gravity critical line in the one-matrix model.
    It has the tricritical point
    $(0.3066\ldots,0.02532\ldots)$ at one end. 
    The dashed curves are unphysical critical lines.}
\end{center}
\end{minipage}
\medskip

\noindent
The dashed curves in Fig.1
are unphysical critical lines%
\footnote{
  The saddle point equation for the eigenvalue distribution 
  \cite{BrItPaZu:planar} generically
  has several branches of solutions.
If we approach the origin within the sheet on which
  the dashed curve lies, the
  distribution does not converge to the Wigner's semi-circle
  distribution. Since the identification of the matrix model
  with gravity is based on the perturbative expansion with
  respect to $g$ around the origin, the dashed curves have nothing
  to do with gravity.}.

We find four solutions to the set of equations 
(\ref{3.10a}--d), as summarized in Table 1:

\noindent
\begin{minipage}{\textwidth}
\[
\begin{array}{|l|c||l|lll|}
\hline
(g_{3*},\ g_{4*}) & \{2-\gamma_0^{(k)}\} & (p,q),\ c_{\rm eff} &
{\cal O}_k & \sim & \int \Psi_{n,m}\ \E^{\beta_{n,m} \varphi} \\
\hline
\hline
( 0.3066\ldots,\ 0.0253\ldots) & 7/3 & (2,5), \ c_{\rm eff}=2/5 &
{\cal O}_1 & \sim & \int \Psi_{1,2}\ \E^{\beta_{1,2} \varphi} \\
\cline{2-2}
\ &  7/2 & \ &
{\cal O}_3 & \sim & \int {\bf 1}\ \E^{\beta_{1,1} \varphi} \\
\hline
(432^{-1/4},\ 0) & 5/2 & (2,3), \ c_{\rm eff}=0 &
{\cal O}_1 & \sim & \int {\bf 1}\ \E^{\beta_{1,1} \varphi} \\
\cline{2-2}
\ &  -6 & \ & \ & \mbox{---} & \ \\
\hline
(0,\ -1/12) & 5/2 & (2,3), \ c_{\rm eff}=0 &
{\cal O}_1 & \sim & \int {\bf 1}\ \E^{\beta_{1,1} \varphi} \\
\cline{2-2}
\ & 5/2 & \ & 
{\cal O}_1 & \sim & \int {\bf 1}\ \E^{\beta_{1,1} \varphi}\\
\hline
(0,\ 0) & -4 & {\rm (Gaussian)} & \ & \mbox{---} & \  \\
\cline{2-2} 
\ & -2 & \ & \ & \mbox{---} & \  \\
\hline
\end{array} 
\]
\begin{center}
{\small 
Table 1: Fixed points of the RG equation
for the one-matrix model and\\
associated critical exponents.
Each fixed point is found to represent a\\ 
continuum
$(p,q)$ gravity through identification of
operators as listed above.}
\end{center}
\end{minipage}

\paragraph{$(2,5)$ fixed point}
We find a fixed point 
$(g_{3*},g_{4*}) = ( 0.3066\ldots,0.0253\ldots)$
on an edge of the pure gravity critical line.
Let us describe how this fixed point is identified with
the (2,5) gravity
(Lee-Yang edge singularity coupled to gravity)
\cite{Staudacher:YangLee}.
By diagonalizing the scaling exponent matrix computed
\beq
G_{ij}=
\lt
\begin{array}{cc}
1.5929\ldots  &  -4.5685\ldots \\
0.1645\ldots  &  -0.1643\ldots 
\end{array}
\rt ,
\eeq
we obtain 
$2/(2-\gamma_0^{(i)})=6/7,\ 4/7$ and
$
V_{ij}=
{\small \lt
\begin{array}{cc}
1  &  -6.2087\ldots \\
1  &  -4.4722\ldots 
\end{array}
\rt }.
$
Therefore the singular part of
the free energy is found to behave
in the neighborhood of the fixed point as
\begin{equation}
  F_{\rm sing}(N,g_3,g_4)
  \sim b\cdot [\delta g_3 -4.4722\ldots \delta g_4 ]^{7/3} 
    + b'\cdot [\delta g_3 -6.2087\ldots \delta g_4 ]^{7/2}
  + \cdots.
  \label{2coupm=3}
\end{equation}
%The position of the fixed point and 
%the transformation matrix 
Values of $g_{i*}$ and $G_{ij}$
can be expressed as roots of algebraic equations, 
though we present them in numerical forms for brevity here.

On the other hand, the content of relevant operators
in the (2,5) gravity consists of 
${\cal O}_1,\ {\cal O}_3$ and ${\cal O}_5$.
Among them ${\cal O}_5$ is a boundary operator
which should be suppressed in our formulation.
According to eq.(\ref{eqn:kthexp}), 
${\cal O}_1\sim \int \Psi_{1,2}\ \E^{\beta_{1,2} \varphi}$
and
${\cal O}_3\sim \int {\bf 1}\ \E^{\beta_{1,1} \varphi}$
acquire the scaling exponents
$2-\gamma_0^{(1)}=7/3$ and 
$2-\gamma_0^{(3)}=7/2$ respectively,
which are identical to those in
eq.(\ref{2coupm=3}).
This completes the identification,
since we generically expect to detect all the
relevant operators (excluding boundary operators)
within the coupling constant space which is
wide enough to realize the fixed point.

\paragraph{(2,3) fixed point I}
We find a fixed point at
$(g_{3*},g_{4*}) = (432^{-1/4},0)$ on the
critical line.
By diagonalizing the scaling exponent matrix computed
\beq
G_{ij}=
\lt
\begin{array}{cc}
4/5 &  17(3^{-3/4}-2\cdot 3^{3/4})/20 \\
0   &  -1/3 
\end{array}
\rt ,
\eeq
the singular part of
the free energy is found to behave as
\begin{equation}
  F_{\rm sing}(N,  g_3 ,  g_4)
  \sim b\cdot 
  \left[ \delta g_3 +
  \left(  \frac{3^{1/4}}{4} - \frac{3^{7/4}}{2}
\right) 
\delta g_4
\right]^{5/2}
+ b'\cdot [\delta g_4]^{-6} + \cdots.
\label{2coupm=2}
\end{equation}
This fixed point is identified with
the (2,3) gravity (pure gravity) whose
sole relevant and non-boundary operator
is the dressed identity
${\cal O}_1\sim \int {\bf 1}\ \E^{\beta_{1,1} \varphi}$
with the susceptibility exponent
$2-\gamma_0^{(1)}=5/2$.
Meanwhile the boundary operator ${\cal O}_3$ is
suppressed as expected.

Here we stress that exponents of irrelevant operators 
${\cal O}_k$ for $k>p+q$, for which
$2-\gamma_0^{(k)}=\gamma_1^{(k)}<0$,
could not be detected correctly within our large-$N$ RG equation.
It is because identification of $N$ with the cutoff
is based upon the possibility of the double scaling
limit (\ref{eqn:doublescalinglimit}) 
where we let $N$ approach infinity 
in correlation to a combination of $g_i$'s
coupled to an operator, sent to a critical value.
Obviously this is possible only for positive $\gamma_1^{(k)}$.
Therefore the magnitude of negative exponents such as $-6$
is an artifact of our large-$N$ RG and untrustable.

\paragraph{(2,3) fixed point II}
At $(g_{3*},g_{4*}) = ( 0, -1/12)$
on the critical line, the $G$-function
turns out to develop singularities of the form
$\beta_{0}(g_i)\sim O\lt\lt V_{ij}\delta g_j\rt^{ 5/6}\rt$,  
$\beta_{1}(g_i)\sim O\lt\lt V_{ij}\delta g_j\rt^{-1/6}\rt$
and so forth.
These singularities originate from the switchover of the 
dominant solution $\bar{\lambda}$ to the saddle point equation
(\ref{9}) or (\ref{eqn:red-spe2}) as we pass over the $g_3=0$ line
\cite{Periwal}.
By replacing the $G$-function in eqs.(\ref{3.10a}--d)
by its regular part $G_{\rm reg}$,
we find that 
$(g_{3*},g_{4*}) = ( 0, -1/12)$
is indeed a fixed point with a scaling exponent matrix
\beq
G_{ij}=
\lt
\begin{array}{cc}
4/5 & 0 \\
0   & 4/5 
\end{array}
\rt .
\eeq
Since eigenvalues of the scaling exponent 
matrix are degenerate,
the transformation matrix $V_{ij}$ in
the singular part of the free energy
\begin{equation}
  F_{\rm sing}(N,  g_3 , g_4)
  \sim b\cdot 
  [V_{1j} \delta g_j]^{5/2} +b' \cdot [V_{2j} \delta g_j]^{5/2} + \cdots.
  \label{2coupm=20}
\end{equation}
remains indeterminate.
This fixed point can also be identified with the (2,3) gravity,
involving doubling of the dressed identity
operator due to the 
$\Z_2$-invariance of the potential%
\footnote{It is well known that the universal part of
the free energy derived from matrix models with
$\Z_2$-symmetric potential
is twice as much as that of matrix models without $\Z_2$-symmetry.}.

\paragraph{Gaussian fixed point}
We find a trivial fixed point at 
$(g_{3*}, g_{4*})=(0, 0)$.
The scaling exponent matrix indicates the naive scaling behavior
\begin{equation}
  F_{\rm sing}(N,  g_3 ,  g_4)
  \sim 
   b \cdot [\delta g_3]^{-4} 
    + b' \cdot [\delta g_4]^{-2} 
   +\cdots.
  \label{2coupm=1}
\end{equation}

\subsection{Fixed points of the two-matrix model}
\label{sec:2mm-crit-pts}
The two-matrix model (\ref{eqn:2mm-cubic-pot})
has been investigated in ref.\cite{Kazakov:Ising},
and is known to possess
the pure gravity critical line parameterized by
\begin{eqnarray}
  g^2 & = & -2c z^2 + z\left\{-c^2 + \left(c + 4 z -
        \sqrt{(1+c)^2 + 8 c z + 16 z^2}\right)^2\right\}, \\
 0 & = & 
 (-1-2c+3c^2 + 24 c z + 48z^2) \nonumber \\
& &\cdot (
1+4c+6c^2+4c^3+c^4+24c z +48 c^2 z +24 c^3 z + 
144 c^2 z^2 + 256c z^3) \nonumber 
\end{eqnarray}
and the tricritical point at one end of the line. 
(See Fig.2).

\noindent
\begin{minipage}{\textwidth}
\begin{center}
    \leavevmode
    \figinclude{200}{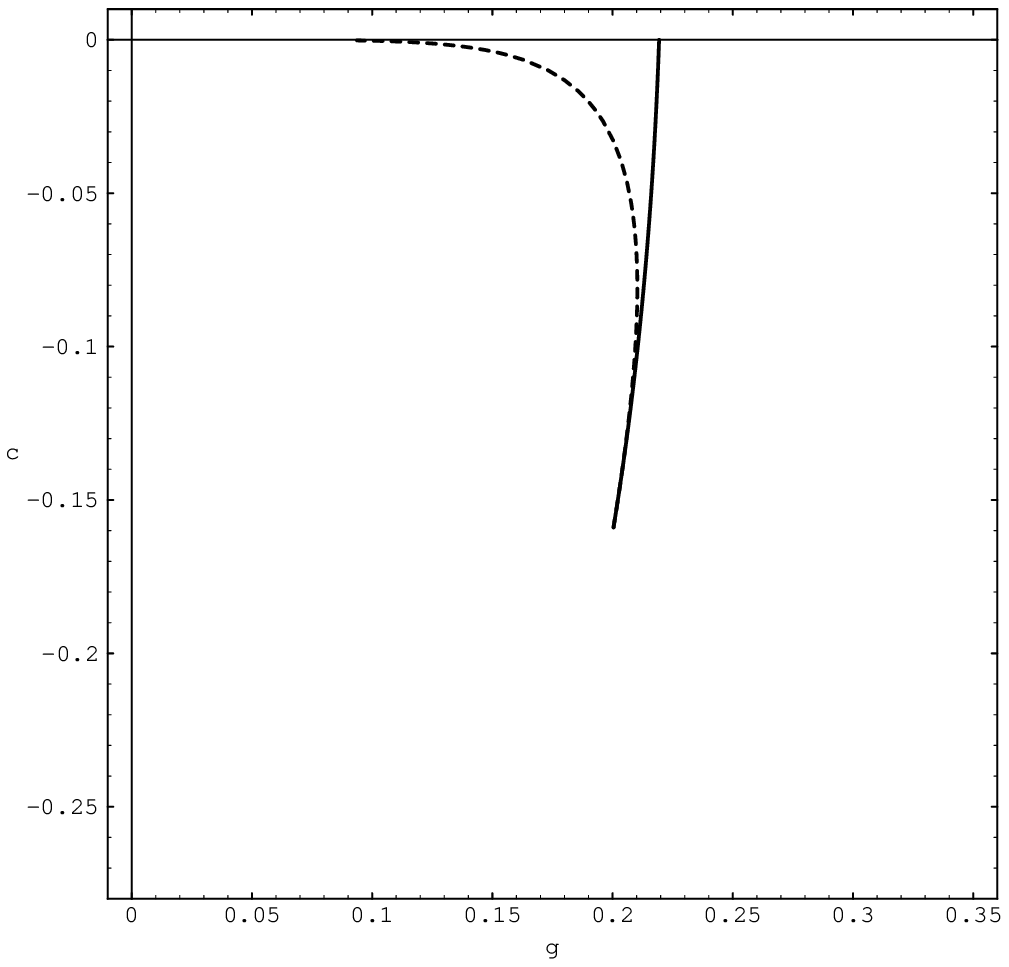}\\
    
{\small Figure 2: 
  The critical line in the two-matrix model with the cubic
    potential.
    The gravity is critical along this line. At
    one end 
    of the line, the Ising
    model becomes critical as well.
   % and the different values of exponents are observed.
    }
 \end{center}
\end{minipage}
\medskip

We find three solutions to the RG equations 
in the physical region of the
coupling constant space $\{(g,c)| g\ge 0, c\le 0\}$,
as summarized in Table 2:

\noindent
\begin{minipage}{\textwidth}
\[
\begin{array}{|l|c||l|lll|}
\hline
(g_{*},\ c_{*}) & \{2-\gamma_0^{(k)}\} & (p,q),\ c_{\rm eff} &
{\cal O}_k & \sim & \int \Psi_{n,m}\ \E^{\beta_{n,m} \varphi} \\
\hline
\hline
(0.2003\ldots,\ 0.1589\ldots)
& 7/3 & (3,4), \ c_{\rm eff}=1/2 &
{\cal O}_1 & \sim & \int {\bf 1}\ \E^{\beta_{1,1} \varphi} \\
\cline{2-2}
\ & 7 & \ &
{\cal O}_5 & \sim & \int \Psi_{1,3}\ \E^{\beta_{1,3} \varphi} \\
\hline
(432^{-1/4},\ 0) & 5/2 & (3,2), \ c_{\rm eff}=0 &
{\cal O}_1 & \sim & \int {\bf 1}\ \E^{\beta_{1,1} \varphi} \\
\cline{2-2}
\ & <0 & \ & \ & \mbox{---} & \ \\
\hline
(0,\ *) & -4 & {\rm (Gaussian)} & \ &  \mbox{---} & \ \\
\cline{2-2} 
\ & 0 & \ & \ & \mbox{---} & \ \\
\hline
\end{array} 
\]
\begin{center}
{\small
Table 2: Fixed points of the RG equation
for the two-matrix model and\\
associated critical exponents.}
\end{center}
\end{minipage}

\paragraph{(3,4) fixed point}
We find a fixed point 
on an edge of the pure gravity critical line,
$(g_*,c_*) =
\left(\left({10(-85+62\sqrt{7})}/{19683}\right)^{1/2},\ 
-(2\sqrt{7}-1)/27\right)$.
By diagonalizing the scaling exponent matrix computed
\beq
G_{ij}=
\lt
\begin{array}{cc}
-0.8391\ldots  &  0.1148\ldots \\
-0.0866\ldots  &  0.3036\ldots 
\end{array}
\rt ,
\eeq
we obtain 
$2/(2-\gamma_0^{(k)})=6/7,\ 2/7$ and
$
V_{ij}=
{\small \lt
\begin{array}{cc}
1  &   -0.2074\ldots \\
1  &    6.3388\ldots 
\end{array}
\rt }.
$
Therefore the singular part of
the free energy is found to behave
as \footnote{
  The logarithmic correction should be present in the free
  energy for the positive integral value of $2-\gamma_0$ to be detected
  in the scaling exponent matrix.
  We have borrowed from the exact solution the fact that the correction takes the
  form $\delta g^7 (\log \delta g)^1$.
} 
\begin{equation}
  F_{\rm sing}(N,g,c) 
  = b \cdot  [\delta g - 0.2074 \cdots \delta c]^{7/3} 
  + b' \cdot [\delta g + 6.3388 \cdots \delta c]^7
        \log [\delta g + 6.3388 \cdots \delta c]%
  + \cdots.
\label{34}
\end{equation}
This fixed point is identified with
the (3,4) gravity (critical Ising model coupled to gravity);
relevant and non-boundary operators of
the (3,4) gravity consist of 
${\cal O}_1\sim \int {\bf 1}\ \E^{\beta_{1,1} \varphi}$, 
${\cal O}_2\sim \int \Psi_{1,2}\ \E^{\beta_{1,2} \varphi}$, and
${\cal O}_5\sim \int \Psi_{1,3}\ \E^{\beta_{1,3} \varphi}$. 
According to eq.(\ref{eqn:kthexp}), 
they acquire the scaling exponents
$2-\gamma_0^{(1)}=7/3$,
$2-\gamma_0^{(2)}=14/5$, and
$2-\gamma_0^{(5)}=7$, respectively.
Since we have imposed the zero magnetic field condition
$g_+=g_-$ at the discrete level, 
the dressed spin operator ${\cal O}_2$ 
should not be detected.
Critical exponents corresponding to
the rest of the operators (the dressed identity and energy)
are indeed detected in eq.(\ref{34}) and this completes the identification.

\paragraph{(3,2) fixed point}
At $(g_*,c_*) = (432^{-1/4},0)$ on the critical line,
the $c$- and $a_c$- derivatives of the $G$-function
turn out to develop singularities.
By replacing the $G$-function in eqs.(\ref{3.10a}--d)
by its regular part $G_{\rm reg}$
as in the (2,3) fixed point II,
we find this point is indeed a fixed point.
Recalling the result of the one-matrix model,
this point obviously corresponds to the 
(twice tensor product of) pure gravity. 

\paragraph{Gaussian fixed line}
We find that all the points on the line $g=0$ are 
fixed points.
The scaling exponent matrix indicates the naive scaling behavior
\begin{equation}
F_{\rm sing}(N,g,c) = 
b \cdot[\delta g]^{-4} +
 b' \cdot [\delta c]^0 +\cdots .
\end{equation}
The existence of the fixed line is a direct consequence of
the separation of the Gaussian two-matrix model 
with arbitrary $c$ into
two Gaussian one-matrix models.

Finally it is important to remark
that the critical behavior of the second term
in eq.(\ref{2coupm=3}), (\ref{2coupm=2}),
or (\ref{34})
is realized at each 
fixed point when one approaches the fixed point along the 
critical line.
This means that, by fine-tuning the most relevant perturbation
by ${\cal O}_1$ to zero, 
we can deform a multi-critical continuum theory while keeping the
gravity critical.
This point will further be discussed in sect.4.2.

%%%%%%% RG Flow  %%%%%%%%%%%%%%%%%%%%%%%%%%
\section{Renormalization group flow}
\setcounter{equation}{0}
\subsection{Linear approximation to the RG equation}
In this section we discuss the structure of the RG flow 
in the coupling space of matrix models. 
For the usual linear RG equation of Callan-Symanzik type
\beq
\left[ N\frac{\pa}{\pa N}+\gamma(g) \right]F(N,g)=
r(g)+\beta(g) \frac{\pa F}{\pa g} ,
\eeq 
the concept of running coupling constant $g(N)$ is 
naturally introduced by
\beq
\int^{g(N)}_{g_0}\frac{dg^{'}}{\beta(g^{'})}
=\log \left({N\over N_0}\right) ,
\eeq
where $g_0=g(N_0)$ and the $\beta$-function plays a role of vector field 
describing RG flow.

We consider, for instance, the RG flow for the one-matrix model
with the cubic coupling described by eq.(\ref{eqn:mat-nrge}) 
with $g_2=1$, $g_3=g$, others$=0$.
Expanding the right hand side into a power series in $\pa F/ \pa g$ 
around $\pa F/ \pa g |_{g=0}=0$, we obtain
\beq
\left[ N \frac{\pa}{\pa N} +2 \right] F(N,g)=\sum_{n=0}^{\infty} \beta_n (g) 
\lt \frac{\pa F}{\pa g} \rt^n +O \lt \frac1N \rt, \label{expandedRGeq}
\eeq
\bea 
\beta_1(g)&\!=\!&-\frac{g}{2} +
3g^3 \int_0^{\frac{1}{1-g^2}}
\frac{dt}{[(1-t)^2+4g^2 t^3 (1-(1-g^2)t)]^{1/2}}, \nonumber \\
\beta_2(g)&\!=\!&-\frac{9g^4}{2(1-g^2)} +
9g^8 \int_0^{\frac{1}{1-g^2}} 
\frac{dt\ t^4}{[(1-t)^2+4g^2 t^3 (1-(1-g^2)t)]^{3/2}}, \nonumber \\
\beta_3(g)&\!=\!&-\frac{9g^5 (3-3g^2+2g^4)}{2(1-g^2)^3}+
54g^{13} \int_0^{\frac{1}{1-g^2}}
\frac{dt\ t^8}{[(1-t)^2+4g^2 t^3 (1-(1-g^2)t)]^{5/2}}, \nonumber \\
&\ & \cdots . \nonumber
\eea
Expansion in powers of $g$ shows 
that $\beta_2(g)\sim -9/2\ g^6+O(g^8)$,
$\beta_3(g)\sim 9/2\ g^9+O(g^{11})$ and so forth. 
Therefore we expect that 
the RG flow should be well approximated by 
$\beta_1(g) (\sim -g/2 - 3g^3 \log g^2)$ 
for the region $\vert g \vert \ll 1 $ where most fixed points lie.
In fact, we have approximate values of fixed points
\beq
\beta_1  (g_*) =0   \mbox{  for  } g_*=0,\ 0.219345 \cdots \ ; \ 
\beta_1^{'}(g_*) =-1/2,  \ 0.780698 \cdots  .
\eeq
By construction $\beta_1(g)$ possesses a zero at 
the Gaussian fixed point
with the exact value of the slope. 
On the other hand the deviations of the location of the nontrivial
fixed point and of the slope from the exact values are 
0.237\% and 2.472\% respectively. 
The small magnitude of the errors is obviously due to the suppression 
factors
multiplying the nonlinear terms in eq.(\ref{expandedRGeq}) 
as stated above. 
We emphasize that the approximate evaluation of $\beta_n$ 
is useful only 
after the reparametrization identities are taken into account,
as is explained in sect.2.1. 

By the same token, 
by linearizing the RG equations
(\ref{eqn:explicit-rge})
around the Gaussian values,
%$g_3=g_4=0$, 
$(\pa F^0/ \pa g_3 ,\ \pa F^0/ \pa g_4) |_{g_3=g_4=0}=(0,1/2)$,
we have also calculated $\beta_1$-functions
$\bigl(G_{,a_3}(g_3,g_4;0,1/2),\ G_{,a_4}(g_3,g_4;0,1/2)\bigr)$ % corr
for the one-matrix model with cubic and quartic couplings. 
Here we exhibit the real part of the $\beta_1$-functions in 
Fig.3.

\noindent
\begin{minipage}{\textwidth}
\begin{center}
    \leavevmode
    \figinclude{250}{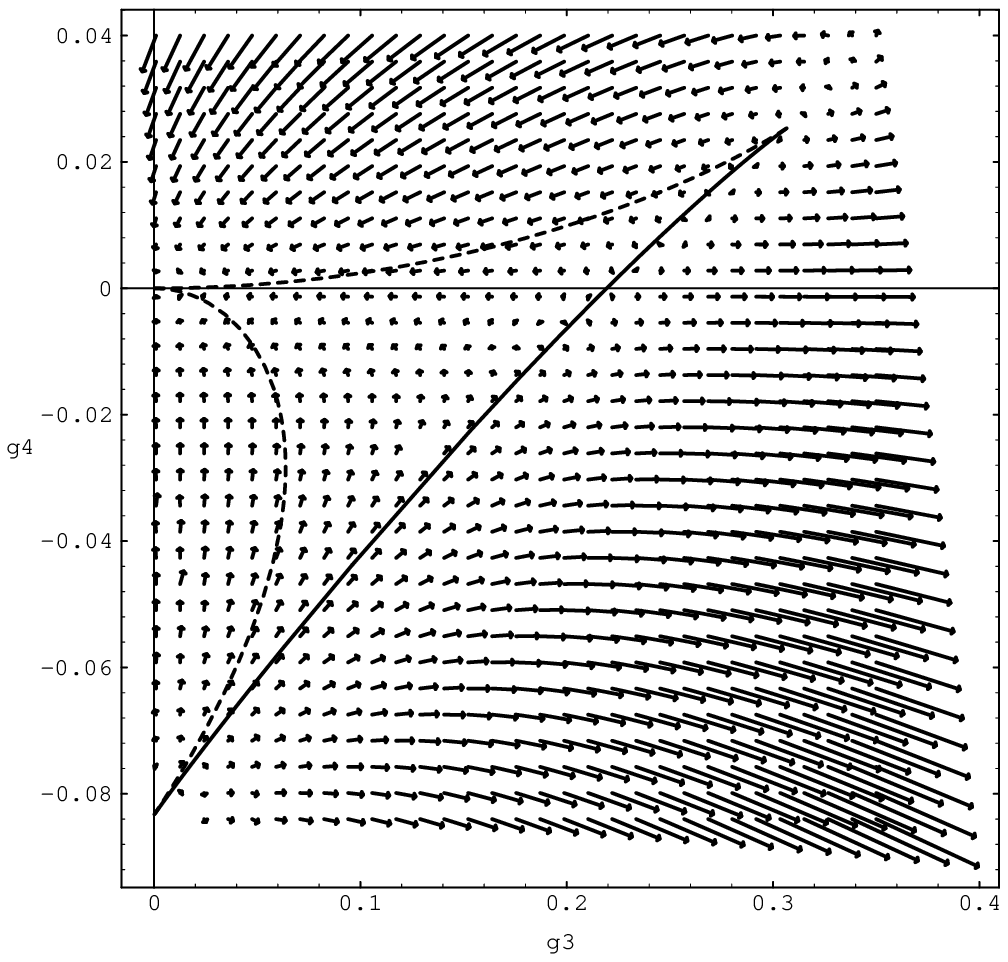}\\
    
{\small Figure 3:
linearized RG flow for the one-matrix model.\\ 
The arrows represent a flow from the UV $(N=\infty)$ to the IR $(N=0)$.}
\end{center}
\end{minipage}
\medskip

\noindent
Similarly the linearization of eq.(\ref{eqn:rgetwomm}) 
around 
%$g=c=0$, 
$(\pa F^0/ \pa g,\ \pa F^0/ \pa c)|_{g=c=0}=(0,0)$ gives 
$\beta_1$-functions 
$\bigl(G_{,a_g}(g,c;0,0),\ G_{,a_c}(g,c;0,0)\bigr)$  % correction
for the two-matrix model (Fig.4).
\medskip

\noindent
\begin{minipage}{\textwidth}
\begin{center}
    \leavevmode
    \figinclude{250}{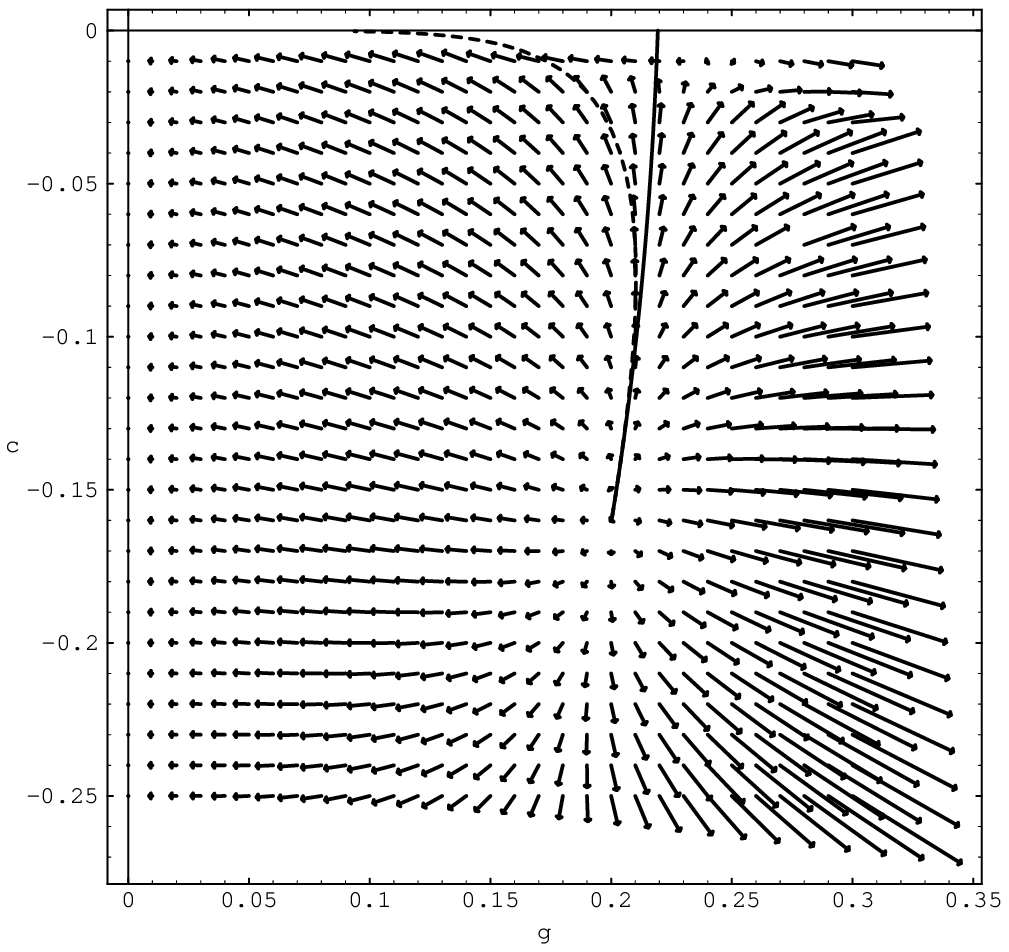}\\ 
 {\small Figure 4: linearized RG flow for the two-matrix model.}
\end{center}
\end{minipage}
\medskip

\noindent
We immediately observe that they reproduce
all the fixed points found in the previous section 
within a very small margin of error. 
Moreover, IR-repulsive and 
attractive directions of the RG flow around fixed points
(which correspond to perturbations by relevant and irrelevant operators,
respectively) are approximated quite well.

\subsection{Global aspects of the RG flow}
The real parts of $\beta_1$-functions plotted in Figs.3 and 4
do not merely reproduce the local properties of the fixed points
but also qualitatively represent 
the global structure of the RG flow.

\noindent {\large \underline{Topology of the RG flow}}

We immediately observe from the figures that
pure gravity critical lines are characterized
as the renormalized trajectories emanating from multi-critical fixed 
points.
By comparing the exponents with the spectrum of exact solutions 
(\ref{eqn:kthexp}),
the trajectory of the one-matrix model in Fig.3 is identified 
with the one corresponding 
to the perturbation of the $(2,5)$ gravity
by its dressed $\Psi_{(1,1)}$ (identity) operator, and that 
of the two-matrix model in Fig.4 
to the perturbation of the $(3,4)$ gravity
by its dressed $\Psi_{(1,3)}$ (energy) operator.
Both of them flow into the pure $(2,3)$ gravity.
These results lead us to propose the following conjectures:
\begin{itemize}
\item
perturbation of an UV theories of two-dimensional gravity 
by its least relevant operators
leads to the IR theory with the neighboring order of criticality; 
\item
the effective central charge\footnote{
The effective central charge in a conformal field 
theory is defined as the measure of 
the rate of divergence of the canonical partition function
of the theory
\[
Z(\beta)={\rm Tr}\ \E^{-\beta H}
\sim \beta^r \exp (\pi c_{\rm eff}/6\beta)
\ \ \ \ \ 
\lt H=2\pi (L_0+\overline{L}_0-c/12) \rt 
\]
as $\beta\rightarrow 0$. 
Since $Z(\beta\rightarrow 0)$ is interpreted as the 
regularized form of the number of states in the theory,
$c_{\rm eff}$ describes the asymptotic growth of density of states.}
$
c_{\rm eff}^{(p,q)} \equiv c^{(p,q)}-24 \Delta_0^{\rm min}= 1-6/pq
$
decreases along the RG flow from
the UV to the IR.
This $c_{\rm eff}$ counts the density of states of the matter sector 
\cite{ItSaZu:nonunitary}.
\end{itemize}
In particular, the
RG trajectory corresponding to the dressed 
$\Psi_{(1,3)}$ perturbation interpolates two neighboring
unitary models. 
This is an intriguing example of
the gravitational counterpart of 
the familiar result on two-dimensional renormalizable field theories
\cite{Zamolodchikov:ctheorem}. 
One may naively anticipate this result from the 
knowledge about the statistical and field theories over fixed 
backgrounds. 
However, we have vanishing total central charge of the system
when we couple the matter with quantum gravity. 
Therefore it is nontrivial to 
establish the concept of `reduction of degrees of 
freedom' along the flow from
the UV to the IR in presence of gravity.

We remark on the RG trajectory between two (2,3) fixed points in Fig.3.
At first sight it is curious that a RG flow apparently exists
to interpolate two distinct fixed points 
(the UV (2,3) fixed point II and the IR (2,3) fixed point I)
corresponding to the same continuum theory.
We interpret this fact as follows.
The one-parameter subspace $(g_3=0,\ g_4)$
is stable under the RG flow,
because our RG transformation maintains the $\Z_2$-invariance
$\phi \rightarrow -\phi$. 
Moreover, the exact solution shows that the singular 
part of the free energy is discontinuous
as we trace it on the critical line from the region $g_3 >0$
to $g_3=0$.
As the $g_3$-derivatives of the free energy is ill-defined,
we can attach no meaning to an RG trajectory connecting 
the $\Z_2$-invariant subspace $g_3=0$ with other regions;
we can make sense only for the eigenvector of the 
scaling exponent matrix and the RG trajectory which are along the $g_4$ 
axis around the (2,3) fixed point II.\\

\noindent {\large \underline{Comparison to the KP flow}}

Let us further contrast these results with the integrable
hierarchical description
of two-dimensional gravity 
\cite{FuKaNa:universal}.
We again recall the fact that $(p,q)$ gravities are
universally described by the $W_p$-constrained,
$p$-reduced KP hierarchy with
a source insertion at $t_{p+q}$ ($t_i$ denotes the time parameter
coupled to the $i$-th order Hamiltonian ${\cal O}_i$
of the hierarchy).
We list the sets of relevant operators (usually referred to as
gravitational primaries)
in $(3,4)$ and $(3,2)$ theories in KdV description of the 
matrix models as well as 
the corresponding operators in the Liouville field theory 
description:

\noindent
\begin{minipage}{\textwidth}
\begin{center}
{\begin{tabular}{crccccccc}
\underline{$(3,4)$}\ \ & KP operators:\ \ &
${\cal O}_1$ &  ${\cal O}_2$ & $\cdot$ & $\lt {\cal O}_4 \rt$ &
${\cal O}_5$ & $\cdot$ & [$ {\cal O}_7$] \\
\ & DDK operators:\ \ &
$\int\Psi_{(1,1)}$ & $\int\Psi_{(1,2)}$ & \ & \ & $\int\Psi_{(1,3)}$ & \ & 
$S^{(3,4)}$ \\
\ & \ & \ & \ & \ & \ & \ & \ & \  \\
\ & \ & \ & \ & \ & \ & $\downarrow$ & \ & \  \\
\ & \ & \ & \ & \ & \ & \ & \ & \  \\
\underline{$(3,2)$}\ \ & KP operators:\ \ &
${\cal O}_1$ &  $\lt {\cal O}_2  \rt$ & $\cdot$ & $\lt {\cal O}_4 
\rt$ & 
[${\cal O}_5$] & \ & \  \\
\ & DDK operators:\ \ &
$\int\Psi_{(1,1)}$ & \ & \ & \ & $S^{(3,2)}$ & \ & \ \\
\end{tabular}}\\
{~}\\
{\small Table 3:~relevant operators of $(3,4)$ and $(3,2)$ theories}\\
\end{center}
\end{minipage}
\medskip

\noindent
In this table supplemented are two types of KP operators:
${\cal O}_{3+q}$ corresponding to the critical action $S^{(3,q)}$ 
parenthesized 
by [\ ], and boundary operators 
${\cal O}_k$, $k=0$ mod $q$ 
parenthesized by (\ ).
It is naturally expected in the KP description 
that the $(3,4)$ gravity
perturbed by ${\cal O}_5$ should flow into the $(3,2)$ gravity;
since ${\cal O}_5$ is relevant, $t_5$ will increase to infinity 
along the RG trajectory against a fixed source value of $t_7$.
Then by rescaling each $t_i$ by $(t_5)^{-i/5}$ 
so as to render $t_5$ finite and $t_7$ infinitesimal 
while keeping the form of the hierarchy, we can describe 
the IR theory by the same 3-reduced KP hierarchy
having a source insertion at $t_5$.
This theory is regarded as the $(3,2)$ gravity.
Our result on the two-matrix model is in accord with this picture.

On the other hand, 
the sets of relevant operators 
of $(2,5)$ and $(2,3)$ theories are:

\noindent
\begin{minipage}{\textwidth}
\begin{center}
{\begin{tabular}{crccccccc}
\underline{$(2,5)$}\ \ & KP operators:\ \ &
${\cal O}_1$ & $\cdot$ &  ${\cal O}_3$ & $\cdot$ &
$\lt {\cal O}_5 \rt$  & $\cdot$ & [$ {\cal O}_7$] \\
\ & DDK operators:\ \ &
$\int \Psi_{(1,2)}$ & \ & $\int \Psi_{(1,1)}$ & \ & \  & \ & 
$S^{(2,5)}$ \\
\ & \ & \ & \ & \ & \ & \ & \ & \  \\
\ & \ & \ & \ & \ & $\searrow$ & \ & \ & \  \\
\ & \ & \ & \ & \ & \ & \ & \ & \  \\
\underline{$(2,3)$}\ \ & KP operators:\ \ &
${\cal O}_1$ & $\cdot$ &  $\lt {\cal O}_3 \rt$ & $\cdot$ &
[$ {\cal O}_5 $]  & \ & \ \\
\ & DDK operators:\ \ &
$\int \Psi_{(1,1)}$ & \ & \ & \ & $S^{(2,3)}$ \  & \ & 
\end{tabular}}\\
{~}\\
{\small Table 4:~relevant operators in $(2,5)$ and $(2,3)$ theories}\\
\end{center}
\end{minipage}
\medskip

It is awkward to interpret the RG flow within the framework of 
the KP hierarchy alone, as we describe below. 
One naively expects that the ${\cal O}_5$ operator perturbation 
would give the $(2,3)$ gravity from the $(2,5)$. 
However, the boundary operators  such as 
${\cal O}_5$ is physically redundant, since it can be 
expressed by means of other primary operators. 
In fact we have found that the RG flow from 
the ${\cal O}_3\ ( \sim \int {\bf 1}\e^{\beta_{1,1} \varphi})$ operator 
perturbation of the $(2,5)$ gravity leads to the $(2,3)$. 
Our large-$N$ RG treatment, which
suppresses appearance of such redundant
operators automatically by the use of 
reparametrization identities, is efficient
indeed in such cases for identifying the RG trajectories
and perturbing operators.

We remind the reader that the `universal' description of 
$(p,*)$ theories by $p$-reduced KP hierarchy is derived
by blowing-up one of the critical regions in the 
bare coupling space $\{g_j\}$.
In other words the KP time parameters $\{t_i\}$
coupled to continuum operators 
are renormalized coupling constants
defined with reference to a specific critical point.
Hence it is a priori not guaranteed that the $p$-reduced 
KP hierarchy can describe an RG flow interpolating between 
two distant critical points $(p,q)$ and $(p,q')$ 
by a limiting procedure with respect to $t_i$.

%%%%%%%  Discussions  %%%%%%%%%%%%%%%%%%%%%%%%%%%%%%%%%%%%%%%%%%%
\section{Discussions}
\setcounter{equation}{0}
In this paper we have constructed the RG equation for
matrix models by regarding $N$ as the cutoff,
and investigated its implication.
First we have performed the RG transformation $N+1\rightarrow N$
to obtain the RG flow in the enlarged coupling space.
We have reduced it
into the original finite dimensional coupling space
by the use of reparametrization identities, at the cost of 
allowing the nonlinearity of the RG equation.
These reparametrization identities are the 
discrete Schwinger-Dyson equations 
for the sphere correlators (\ref{eqn:sd-eq}) 
and (\ref{eqn:sdtwomat}). 
It is crucial for this procedure that these equations 
close algebraically to determine the required resolvent.
The RG equation is found to give the critical points
of the exact solution. It also 
exactly reproduces the spectrum of 
relevant operators of the Liouville gravity at its fixed points.
After confirming the adequacy of linear approximation,
we have investigated the RG flow in the coupling space.
{From} these examples, we conjecture that a kind of the gravitational 
analogue of Zamolodchikov's $c$-theorem holds in accordance with 
some previous expectations \cite{KuSa:gravctheorem}.

One of the characteristics of our treatment is that
it suppresses the boundary operators which inevitably
arise from the usual KP description of quantum gravity
derived from matrix models,
but are absent in the BRST analysis of Lian and Zuckerman's 
\cite{LiZu:BRST}.
A direct consequence of this is that
we do not need them to unambiguously determine 
the RG flow.
Further, if applied to multi-matrix models it
can in principle describe the RG flow which changes both $p$ and $q$ 
of the $(p,q)$ theories 
that is impossible to realize in the KP description.
Work in this direction is in progress.

In order to establish the `gravitational $c$-theorem',
it is necessary to construct 
a counterpart of the $c$-function
in the framework of matrix models.
The $c$-function is a potential function for 
the $\beta$-function, which should monotonically decrease
along the RG trajectory from the UV extremum to the IR one. 
It might exist, since the $G$-function itself
plays the role of the $c$-functions in nonlinear RG equations.

As originally motivated Br\'{e}zin and Zinn-Justin,
another intriguing application of our method is 
to solve matrix models 
which do not allow angular integration. 
Such models include candidates for
a system of $c>1$ matter coupled to gravity. 
For instance, the $n$-Ising model 
over randomly triangulated surfaces can be described by 
a matrix model with $2^n$-plets of $N\times N$ matrices 
$\phi_1, \cdots,\phi_{2^n}$ as:
\bea
Z(N,g,\beta)&\!\!=&\!\!\int \prod_{\bt} d^{N^2} \phi_{\bt} \ 
\exp \left\{ -N \tr V(\phi_1, \cdots, \phi_{2^n}) \right\}, 
\nonumber \\
V(\phi_1, \cdots, \phi_{2^n})
&\!\!=&\!\!
 \hf \sum_{\bt,\bt'} \phi_{\bt}
\Delta_{\bt \bt'} \phi_{\bt'} +\frac{g}{3} \sum_{\bt} \phi^3_{\bt}
\nonumber \\
\bt 
&\!\!=&\!\!
(\sigma_1,\sigma_2,\cdots,\sigma_n), 
\qquad  \sigma_i=\pm 1,
 \nonumber \\
\lt \Delta^{-1} \rt_{\bt \bt'}
&\!\!=&\!\!
\e^{\beta \bt \cdot \bt'}, 
\qquad \bp \equiv 
 \left(
\begin{array}{ccc}
 \mbox{\PHIONE}&      &\ZERO        \\
               &\ddots&             \\
   \ZERO       &      &\mbox{\PHITWO}
\end{array}
\right) .
\eea
It is straightforward to write down the RG and saddle point equations,
\begin{eqnarray}
&\!\!\!&\!\!\!\left[ N\frac{\partial}{\partial N} +2 \right] F(N,g,\beta)
= \la {1 \over N} \tr V(\phi_1, \cdots, \phi_{2^n}) \ra 
+ {1 \over \cosh^n \beta}{ \la \as \ra^2 \over 2} 
+ 2^n {g \over 3} \la \as \ra^3 \nonumber \\     
&\!\!\!&\!\!\! \ 
+\la \frac{1}{N} {\rm Tr} \log \left( {\bf 1} \Delta_{\bt \bt'} 
+g\left(\la \as \ra{\bf 1} \delta_{\bt \bt'} 
+ \phi_\tau  \delta_{\bt \bt'} \right)
\right) \ra
 -2^{n-1}+O \left( \frac{1}{N} \right) ,
\end{eqnarray}
\begin{eqnarray}
{1 \over (2\cosh \beta)^n}
 \la \as \ra
+g \la \as \ra^2 
+g \la \frac{1}{N}
{\rm tr} 
\left({1 \over {\bf 1}\Delta+g(\la \as \ra {\bf 1}\delta
+\mbox{\boldmath{$\phi$}})}
\right)_{\bt \bt}  \ra
=0,
\end{eqnarray}
where the trace over the $(2^n\cdot N) \times (2^n\cdot N)$ matrix is 
denoted by Tr, and
the trace over $N \times N$ denoted by tr. 
We have assumed $\alpha_{s,\tau}$ does not depend on $\tau$
in the same way as in the analysis of the two-matrix model.
Thus the problem reduces to the computation of the resolvent
\beq
W(z)=
\left\langle \frac1N {\rm Tr} \frac{1}{ z {\bf 1} \delta
 +({\bf 1} \Delta +g \bp )  } \right\rangle 
\eeq
by using the reparametrization identities.
So far we have not yet seen whether the resolvent can be 
determined by a closed set of Schwinger-Dyson equations or not.
Even if we do not obtain a finitely closed set, however, 
we can exploit
available approximation schemes, such as the linearization
of the RG equation.

(Non-polynomial) nonlinearity is inevitable in constructing
an RG equation obeyed by the free energy $F(N,g)$.
Consequently nonlinearity makes the identification of fixed points and 
exponents complicated, and makes the concept of running 
coupling constant unclear.
While we are writing the manuscript we noticed two recent papers 
which may be relevant to this point.
In ref.\cite{Johnston:penner} the free energy of the
Penner matrix model 
is shown to obey a linear RG equation.
It is of great interest to examine whether the linearity
of the RG equation is the peculiarity of the Penner model
or a generic feature in a certain class of matrix models.
A more recent paper \cite{Hikami:perturbative} 
suggests an interesting possibility for constructing a linear RG
equation 
by using a mapping of matrix models to $O(N^2)$-vector models.
It was also proposed to consider the RG equation for 
the first derivative of the free energy $\pa F / \pa g$. 
Although generalizations of this mapping to the subleading orders
in $N$ are hard to provide, 
so far we have observed that a certain simplification occurs
when we perform an RG transformation for $\pa F / \pa g$;
there appears neither the $\tr \log$-term present in
eq.(\ref{eqn:sp-differential}), nor the integration term 
in eq.(\ref{eqn:mat-nrge}).

We hope to report on these topics in a future publication.\\

\noindent {\Large \bf Acknowledgments}\\

\noindent 
The authors thank 
M.~Fukuma for helpful discussions,
S.~Hikami for discussions and explaining his work, and 
P.~Crehan for carefully reading the manuscript. 
This work is supported in part by
Grant-in-Aid for Scientific Research (S.H.) and
(N.S., No.05640334), and Grant-in-Aid for Scientific Research
for Priority Areas (N.S., No.05230019) {}from the Ministry of
Education, Science and Culture.

\renewcommand{\thesection}{A}
%%%%% Loop Eq for 2MM %%%%%%%%%%%%%%%%%%%%
\section{Loop equation for the two-matrix model}
\setcounter{equation}{0}
\renewcommand{\theequation}{A.\arabic{equation}}
In this appendix we exhibit the explicit form of the one-point
function of the resolvent operator $ \awa{0} $ in 
eq.(\ref{eqn:resolventtwomm}) 
by using the reparametrization invariance of the partition
function (\ref{eqn:2mm-cubic-pot}).
We will consider the following reparametrizations:
\renewcommand{\theequation}{A.\arabic{equation}\alph{subeqn}}
\setcounter{subeqn}{1}
\begin{eqnarray}
  \bp' &\!\!=\!\!& \bp 
  + \epsilon \sum_{a=\pm}
  P_a \left( z \bs_0 + \frac cg \bs_1 +  \bp\right)^{-1} P_a,
  \label{eqn:2mm-reparam-1} \\
\addtocounter{equation}{-1}
\addtocounter{subeqn}{1}
  \bp' &\!\!=\!\!& \bp 
  + \epsilon \sum_{a=\pm}
  P_a  \left\{ \bs_1, 
  \left( z \bs_0 + \frac cg \bs_1 +  \bp \right)^{-1} \right\} P_a,
  \label{eqn:2mm-reparam-2} \\
\addtocounter{equation}{-1}
\addtocounter{subeqn}{1}
  \bp' &\!\!=\!\!& \bp \label{eqn:2mm-reparam-4}\\ 
  &\!\!+ \!\!& \epsilon \sum_{a=\pm}
  P_a 
  \lt \bp\bs_1 \left( z \bs_0 + \frac cg \bs_1 +  \bp \right)^{-1} +
  \left( z \bs_0 + \frac cg \bs_1 +  \bp \right)^{-1} \bs_1 \bp \rt P_a,
   \nonumber \\
\addtocounter{equation}{-1}
\addtocounter{subeqn}{1}
 \bp' &\!\!=\!\!& \bp 
  + \epsilon \sum_{a=\pm}
  P_a 
  \left\{\bs_1\bp\bs_1, \left( z \bs_0+ \frac cg \bs_1 +  \bp \right)^{-1} 
\right\} 
  P_a,                 \label{eqn:2mm-reparam-5} 
\end{eqnarray}
These reparametrizations maintain $\phi_\pm$ hermitian.
Invariance under these  reparametrizations induces the following
reparametrization identities
\setcounter{subeqn}{1}
\begin{eqnarray}
0 & = & 
  \frac 12 \awa{0}^2 
  + \left( gz^2 - z + \frac{c^2}{g}\right)\awa{0}
  + \left(2 c z - \frac c g\right) \awa{1} \nonumber \\
& &
  + c \awa{11} + g \left\langle  \oon \Tr \bp \right\rangle + 2 - 2gz, \\
\addtocounter{equation}{-1}
\addtocounter{subeqn}{1}
0&= & 
  \frac 12 \awa{0} \awa{1}
  + \left( c z - \frac cg - \frac{c^2}{g}\right) \awa{0}
  + ( gz^2 - cz - z)  \awa{1} \nonumber \\  & & -c  \awa{11},\\
\addtocounter{equation}{-1}
\addtocounter{subeqn}{1}
0& = &
  -\frac{c}{2g} \awa{0}^2  - \frac 12 z\awa{0}\awa{1} 
   + \left(\frac {cz}{g} - c z^2\right) \awa{0} \nonumber \\ % correction
& & + (-1 + z^2 - g z^3) \awa{1} 
   + \left(- \frac c g  + c z \right) \awa{11}  - c \awa{101}  \\
& & + c \awa{111}, \nonumber\\
\addtocounter{equation}{-1}
\addtocounter{subeqn}{1}
0 & = &  \frac 12 \awa{11}\awa{0}
  + \frac{c^2}{g^2} ( 1- gz) \awa{0} 
  + \frac cg z ( 1 -g z) \awa{1} \nonumber\\
& &  - z ( 1 - gz) \awa{11}  + c \awa{101}  - c \awa{111}
  + (1 - gz)  \left\langle \frac 1N \Tr \bp \right\rangle  \nonumber\\
& & + g \left\langle \oon \Tr \bs_1 \bp \bs_1 \bp \right\rangle,
\end{eqnarray}
\renewcommand{\theequation}{A.\arabic{equation}}%
where
\begin{equation}
\hat{W}_{j_1 j_2 \cdots j_k}(z)=
\frac 1N \Tr \bs_{j_1} \bp \bs_{j_2} \bp \cdots \bp 
\bs_{j_k} 
\left(z \bs_0 + \frac cg \bs_1+\bp\right)^{-1} .
\end{equation}

After some computations, we obtain a quartic equation
obeyed by $\awa{0}$:
\begin{eqnarray}
0
& =  & 
\left( \frac 12 \awa{0} - \frac cg -z + c z + g z^2 \right)
\nonumber \\
&\cdot& \left[-\frac{c}{2g} ( 1 + 2 c - g z) \awa{0}^2   +
 \left(2 \cdot \frac{c^2}{g^2} + \frac{c}{g^2}  - \frac cg z \right)  
( c + g z ) ( 1 -g z) \awa{0}\right.  \nonumber \\     
&+&   \left.    c ( 1 -g z) \left\langle \frac 1N \Tr \bp \right\rangle 
     + 2 g c \la \frac 1N \Tr \bs_1 \bp \bs_1 \bp
     \ra  \right] \nonumber\\
& -&  \left\{\frac 14 \awa{0}^2 +
    \left[ -c z - \left(\frac{c}{2g} + z\right)(1-g z)\right] \awa{0}
    -c  \right.\label{2mmloopquart}
\\
& +& \left.\frac zg ( 1 + 2 c - g z) ( c + g z)(1 -g z)\right\}  
                         \nonumber\\
& \cdot&   \left\{\frac 12 \awa{0}^2 + 
 \left(-\frac cg - z + c z  + g z^2 \right)\awa{0} +
 g\left\langle  \frac 1N \Tr \bp \right\rangle + 2 -2g z \right\}
\nonumber
\end{eqnarray}
There appear one-point functions 
$\langle \tr \phip \phim \rangle=
(1/2)\la \Tr \bs_1 \bp \bs_1 \bp \ra$ and
$\langle \tr (\phip + \phim) \rangle =
\left\langle \Tr \bp \right\rangle$
in the above expression. 
We express the latter in terms of 
$\la \tr \phip \phim \ra$ and $\la \tr (\phip^3 + \phim^3) \ra$
by solving the first two equations in (\ref{eqn:sdtwomat}),
\renewcommand{\theequation}{A.\arabic{equation}\alph{subeqn}}
\setcounter{subeqn}{1}
\begin{eqnarray}
0&=&\la \frac 1N \tr \phipm \ra
+g \la \frac 1N \tr \phi_{\pm}^2 \ra
+c \la \frac 1N \tr \phi_{\mp}   \ra,\\
\addtocounter{equation}{-1}
\addtocounter{subeqn}{1}
1&=& \la \frac 1N \tr \phi_{\pm}^2 \ra
+g \la \frac 1N \tr \phi_{\pm}^3 \ra
+c \la \frac 1N \tr \phip \phim  \ra.
\end{eqnarray}
\renewcommand{\theequation}{A.\arabic{equation}}%
Promoting
$\partial F/\partial g =\langle (1/3N) \tr (\phip^3+\phim^3) \rangle$ and 
$\partial F/\partial c = \langle (1/N) \tr \phip \phim \rangle$ to independent variables,
eq.(\ref{2mmloopquart}) with this replacement determines 
the one-point function of the resolvent
\begin{equation}
  \langle \hat{W}_0(z) \rangle = 
W_0\left(z;g,c;
\frac{\partial F}{\partial g},
\frac{\partial F}{\partial c}
\right) .
\end{equation}

\ifpublisher\newpage\else\end{document}\fi
\pagestyle{empty}
{\Large\bf Figure captions}

\noindent
Figure 1:

The pure gravity critical line in the one-matrix model.
    It has the tricritical point
    $(0.3066\ldots,0.02532\ldots)$ at one end. 
    The dashed curves are unphysical critical lines.
\medskip

\noindent
Figure 2: 

  The critical line in the two-matrix model with the cubic
    potential.
    The gravity is critical along this line. At
    one end 
    of the line, the Ising
    model becomes critical as well 
    and the different values of exponents are observed.
\medskip

\noindent
Figure 3:

linearized RG flow for the one-matrix model.\\ 
The arrows represent a flow from the UV $(N=\infty)$ to the IR $(N=0)$.
\medskip

\noindent
Figure 4: 

linearized RG flow for the two-matrix model.
\newpage
\noindent
\begin{center}
    \leavevmode
    \figinclude{450}{mat-critline.eps}
\end{center}

\vspace*{\fill}
Fig.1\newpage
\begin{center}
    \leavevmode
    \figinclude{450}{2mm-critline-pos.eps}
\end{center}

\vspace*{\fill}
Fig.2
\newpage
\begin{center}
    \leavevmode
    \figinclude{450}{mat-flow.eps}
\end{center}

\vspace*{\fill}
Fig.3
\newpage
\begin{center}
    \leavevmode
    \figinclude{450}{2mm-flow-pos.eps}
\end{center}

\vspace*{\fill}
Fig.4
\end{document}